\documentclass[11pt]{iopart}
\usepackage{iopams}
\usepackage{epsfig}
\usepackage{amssymb}
\usepackage{color}
\usepackage{graphicx}
\usepackage{epstopdf}
\usepackage{float}
\usepackage{color}
\usepackage{siunitx}
\usepackage{caption}
\usepackage{subcaption}
\usepackage{graphicx}
\usepackage{comment}
\usepackage[margin=2.6cm]{geometry}
\usepackage[breaklinks=true,colorlinks=true,linkcolor=blue,urlcolor=blue,citecolor=red]{hyperref}  
\begin{document}

\title{Neutron Shell Closure at N=32 and N=40 in Ar and Ca Isotopes}
\author{M. EL ADRI and M. OULNE}
\address{High Energy Physics and Astrophysics Laboratory, Department of Physics, Faculty of Sciences Semlalia, Cadi Ayyad University P.O.B 2390, Marrakesh 40000, Morocco.}
\eads{\mailto{\textcolor[rgb]{0,0.35,1.0}{eladri.mohamed@gmail.com}
	}, \,\,\,\,\, 
\mailto{\textcolor[rgb]{0,0.35,1.0}{oulne@uca.ma}(Corresponding author)}}

	\begin{abstract}
		In this paper, we  investigate features of the ground state of some nuclei far from the stability for   isotope chains with proton numbers Z=18 and 20. Our aim is to predict the eventual existence of magic numbers in these exotic nuclei. For this purpose, we use two methods: the non relativistic Hartree-Fock-Bogoliubov (HFB) approach based on SLy4 Skyrme functional and the relativistic (so-called covariant) density functional theory (CDFT) by using the DD-ME2 force parametrization.
		We compare our results with the available experimental data and with the predictions of other models such as Finite Range Droplet Model (FRDM). Our	present investigation predicts that N=32 and N=40 are magic numbers for Ar and Ca isotopes.
	\end{abstract}

\vspace{1pc}
\noindent{\it Keywords}: {{Hartree-Fock-Bogoliubov method (HFB); Relativistic density functional theory (CDFT);  New magic numbers; Nuclei far from the stability; Ca and Ar isotopes; neutron paring gap; one- and two-neutron separation energies.}
	
	\section{Introduction}
	Nowadays, nuclear structure continues to be  an active area of research in nuclear physics. Previously, the study of the structure of the Nucleus has rested on stable nuclei or those located near the valley of  stability. However, over the years, the field of investigation of the structure of the nucleus has been extended to exotic nuclei. Nuclear models, which are essentially based on nuclei close to stability, diverge as stability limits are approached. This gave rise to new theories that are developed to describe stable nuclei and applied for some exotic ones.
	
	Generally, the new nuclear theories can be grouped into two different approaches: ab-initio methods \cite{Pud,Pudl} and relativistic\cite{Bender,Serot,Ring,Poschl} or non-relativistic\cite{Dobaczewsk1984,Davies,Negele} mean field theories. In the former case, thanks to the advances of the computing facilities, we can study with success the light nuclei. In the past 15 years or so there has been a lot of effort invested in ab-initio methods (e.g., QMC methods, coupled cluster methods, no-core shell model) and much progress is being made toward the description of medium-mass nuclei \cite{Carlson,Hagen,Barrett}. However, their use for the description of heavy nuclei is still limited because of the complexity of the calculations. Currently , there are methods developed to bridge the ab-initio and DFT approaches\cite{Drut,Grasso}. In the latter case, mean field theories are less complicated and they described nuclear properties successfully. For these reasons, they are mostly used to study nuclear structure.
	
	In the present work, we concentrate on  mean field theories by choosing the non-relativistic Skyrme-Hartree-Fock-Bogliubov method and covariant density functional theory (CDFT) in order to predict and discuss the signature of  new magic numbers in even-even $_{20}Ca$ and $_{18}Ar$ nuclei located far from the stability. 
	
	As it is known, the magicity in nuclei can be changed locally in those which are far away from the stability line. Therefore, the known magic numbers  for nuclei located in the  stability valley or very close to it can disappear and new ones can appear instead. Our work is a part of this current problematic inasmuch as  it concerns the discovery of new magic numbers in the exotic region. 
	Recently, several reviews have studied the emergence of new shell magic numbers, for example N=16 in Z=8 \cite{Ozawa} and N=32 in Z=20 \cite{tendeur,Huck}. However, the other ones are quenching like N=20 in magnesium and neon \cite{Kanungo,Grawe}, and N=28 in sulfur and silicon \cite{Jurado}. Also, a discussion with experimental techniques on magic numbers can be found in the paper by Nakamura et al. \cite{Nakamura}. In the same context, in Refs. \cite{Sorlin,Otsuka} the authors emphasized the role of nuclear forces on shell evolution studies. 
	
	The paper is organized as follows: a short description of the approaches we have used for our calculations and the methods to solve them are given in sections 2 and 3. The details of the input and  the interactions used in calculations are provided in section 4. Our results are presented and discussed in section 5. Finally we give our conclusion in section 6.
	
	\section{Hartree-Fock-Bogoliubov Method}
	In the Hartree-Fock-Bogoliubov approximation, the two-body Hamiltonian is essentially reduced to a sum of two terms: the kinetic energy $t_{i j}$ and the  anti-symmetric two-body interaction matrix-elements $\bar{\nu}_{i j k l}$. So, in second quantization this Hamiltonian takes the form\cite{Ringp}:
	\begin{equation}
	H=\sum_{i j} t_{i j} c_{i}^\dagger c_{j} + \frac{1}{4} \sum_{i j k l} \bar{\nu}_{i j k l} c_{i}^\dagger c_{j}^\dagger c_{l} c_{k}
	\label{eq1}
	\end{equation}
	with $c_{i}^\dagger$ and $c_{i}$ are single-particle creation and annihilation operators, respectively, and $\bar{\nu}_{i j k l}=\langle i, j | V | k, l \rangle-\langle i, j | V | l, k \rangle$ are anti-symmetrized two-body interaction matrix-elements.
	The basic idea in the HFB method is to define ground state of the many-body system as a vacuum with respect to quasi-particles\cite{Landau, Migdal}:
	\begin{equation}
	\beta_k|\Phi\rangle=0
	\end{equation}
	where  $|\Phi\rangle$ is  the ground-state wave function, and $\beta$ and $\beta^\dagger$ are the quasi-particle operators which  can be obtained from the particle operators $c_{i}$ and $c_{i}^\dagger$  by using the general linear Bogoliubov transformation\cite{Ringp}:
	\begin{eqnarray}
	\left(\begin{array}{c} \beta  \\
	\beta^\dagger \end{array}\right ) =\left(\begin{array}{cc} U^\dagger & V^\dagger \\ V^{T} & U^{T} \end{array}\right) \left(\begin{array}{c} c \\ c^\dagger \end{array} \right)
	\end{eqnarray}
	The basic building blocks of the theory are, namely: the one-body density matrix which is given by:
	\begin{equation}\label{rho}
	\rho_{i j}=\langle\Phi|c_{j}^\dagger c_{i}|\Phi\rangle=(V^{*}V^{T})_{ij}
	\end{equation}
	and the pairing tensor defined as: 
	\begin{equation}\label{kappa}
	\kappa_{i j}=\langle\Phi|c_{j}c_{i}|\Phi\rangle=(V^{*}U^{T})_{i j}\
	\end{equation}
	By applying the variational principle, the expectation value of the Hamiltonian (\ref{eq1}) is expressed as an energy functional
	\begin{equation}
	E[\rho,\kappa]=\frac{\langle\Phi|H|\Phi\rangle}{\langle\Phi|\Phi\rangle}=
	\textrm{Tr}[(t+\frac{1}{2}\Gamma)\rho]-\frac{1}{2}\textrm{Tr}[\Delta\kappa^{*}]
	\label{energyfunctional}
	\end{equation}
	where\\
	\begin{equation}
	\Gamma_{i k}=\sum_{j l}\bar\upsilon_{i j k l}\rho_{l j}\,,~~~~~~~~~~         \Delta_{i j}=\frac{1}{2}\sum_{k l}\bar\upsilon_{i j k l}\kappa_{k l}\,.
	\end{equation}
	The variation of the energy (\ref{energyfunctional}) with respect to $\rho$ and $\kappa$ leads to the HFB equations:
	\begin{eqnarray}\label{HFBeq}
	\left(\begin{array}{cc} h-\lambda & \Delta \\
	-\Delta^* & -(h+\lambda)^*
	\end{array} \right)\left(\begin{array}{c} U \\ V \end{array} \right) =E\left(\begin{array}{c} U \\ V \end{array}\right), \, \label{eqhfb}
	\end{eqnarray}
	where $ h=t+\Gamma$ is the mean field Hamiltonian, and $\Delta$ denotes the pairing potential. Later on, the Lagrange multiplier  $\lambda$  will turn out to be the Fermi energy of the system.
	In practice, it is convenient to transform the standard HFB equations into a coordinate space representation and solve the resulting differential equations on a lattice. We can then use the Skyrme forces\cite{chabanat} to conveniently simplify further the HFB equations.  So, in coordinate space the HFB energy (\ref{energyfunctional}) has the form of local energy density functional:
	
	\begin{equation}
	E[\rho,\tilde{\rho}]=\int d^{3}r\textrm{H}(\textbf{r}),
	\label{skyrmeefunctional}
	\end{equation}
	The Hamiltonian $\textrm{H}(r)$ is composed of several terms
	\begin{equation}
	\textrm{H}=K+H_{0}+H_{LS}+H_{C}
	\end{equation}
	The first term is the kinetic energy, the second one corresponds to the density-dependent and the third term represents the finite-range spin-orbit. Finally, $H_{C}$ denotes the coulomb term.
	The variation of the energy (\ref{skyrmeefunctional}) according to the particle local
	density $\rho$ and pairing local density $\tilde{\rho}$ results in Skyrme HFB equations:
	\begin{eqnarray}
		\sum_{\sigma'}\left(\begin{array}{cc} h(\textbf{r} \sigma,\textbf{r} \sigma') & \tilde{h}(\textbf{r} \sigma,\textbf{r} \sigma') \\
	\tilde{h}(\textbf{r} \sigma,\textbf{r} \sigma') & -h(\textbf{r} \sigma,\textbf{r} \sigma')
	\end{array} \right)\left(\begin{array}{c} U(E,\textbf{r}\sigma') \\
	V(E,\textbf{r}\sigma')\end{array} \right)=
	\end{eqnarray}
 \begin{eqnarray}\nonumber
   	\qquad 	\qquad  \qquad	\qquad 	\qquad 	\qquad 	\left(\begin{array}{cc} E+\lambda & 0\\
	0 & E-\lambda)\end{array}\right)\left(\begin{array}{c} U(E,\textbf{r}\sigma) \\
	V(E,\textbf{r}\sigma) \end{array}\right) \,
	\label{shfb}
	\end{eqnarray}
	The local fields $h(\textbf{r} \sigma,\textbf{r} \sigma')$ and $\tilde{h}(\textbf{r} \sigma,\textbf{r} \sigma')$ can be easily calculated in coordinate space (See Refs \cite{Ringp,Stoitsov,Greiner} for more details).
	
	\section{Covariant density functional theory}
	The Covariant density functional theory (CDFT) is a modern theoretical tool for the description of the ground state properties of nuclei. There are three classes of covariant density functional models: the nonlinear meson-nucleon coupling model (NL)\cite{Abusara}, the density-dependent meson-exchange model (DD-ME)\cite{Lalazissis} and the density-dependent point-coupling model (DD-PC)\cite{Niksic}. In the framework of this paper we have used DD-ME class with finite masses leading to finite-range interactions. The starting point of CDFT for these models is a standard Lagrangian density\cite{Niksik2014}
	\begin{eqnarray}
	\mathcal{L}  &  =\bar{\psi}\left[
	\gamma(i\partial-g_{\omega}\omega-g_{\rho
	}\vec{\rho}\,\vec{\tau}-eA)-m-g_{\sigma}\sigma\right]  \psi\nonumber\\
	&  +\frac{1}{2}(\partial\sigma)^{2}-\frac{1}{2}m_{\sigma}^{2}\sigma^{2}%
	-\frac{1}{4}\Omega_{\mu\nu}\Omega^{\mu\nu}+\frac{1}{2}m_{\omega}^{2}\omega
	^{2}\label{lagrangian}\\
	&  -\frac{1}{4}{\vec{R}}_{\mu\nu}{\vec{R}}^{\mu\nu}+\frac{1}{2}m_{\rho}%
	^{2}\vec{\rho}^{\,2}-\frac{1}{4}F_{\mu\nu}F^{\mu\nu}\nonumber
	\label{lagrangian}
	%\end{align}
	\end{eqnarray}
	with $m_{\sigma}$, $m_{\omega}$, $m_{\delta}$ and $m_{\rho}$ are meson masses,  $g_{\sigma}$, $g_{\omega}$, $g_{\delta}$ and $g_{\rho}$ are the coupling constants, and $\Omega_{\mu\nu}$, ${\vec{R}}^{\mu\nu}$, $F_{\mu\nu}$ are fields tensors. e corresponds to the proton's charge. It vanishes for neutron.\\
	The Hamiltonian density reads\cite{Ring}:
	\begin{eqnarray}
		\label{Eq:Ham}
		{\cal H}(\mathbf{r}) &= \sum_i^A\psi_i^\dagger \left(\mathbf{\alpha}\mathbf{p} + \beta m  \right)\psi_i\nonumber\\
		&+\frac{1}{2}\left[(\mathbf{\nabla} \sigma)^2 + m_\sigma^2\sigma^2  \right]%
		-\frac{1}{2}\left[(\mathbf{\nabla}\omega)^2 + m_\omega^2\omega^2  \right]\nonumber\\
		&-\frac{1}{2}\left[(\mathbf{\nabla}\rho)^2 + m_\rho^2\rho^2  \right]%
		-\frac{1}{2}(\mathbf{\nabla} A)^2\nonumber\\
		&+\left[ g_\sigma \rho_s \sigma + g_\omega j_\mu\omega^\mu+g_\rho \vec{j}_\mu\cdot\vec{\rho}^\mu
		+ej_{p\mu}A^\mu  \right].
	\end{eqnarray}
	The most successful EDF (Energy Density Functionals) originates from the Relativistic Hartree-Fock-Bogoliubov (RHFB) model in which p-h and p-p channels are treated simultaneously in a self-consistent manner. In RHFB model, the CDF energy functional is determined by the expectation of the system Hamiltonian ${\cal{H}}$ with respect to the ground-state wave function $|\Phi\rangle$:
	\begin{eqnarray}\label{Energy}
	E=\langle\Phi|{\cal{H}}|\Phi\rangle
	\end{eqnarray}
	The variation of the energy functional given by eq (\ref{Energy}) with respect to Dirac spinor $\psi(r)$ leads to RHB (Relativistic Hartree-Bogoliubov) energy density functional as:
	\begin{equation}
	\label{Eq:EDFP}
	E_{RHB}[{\rho},{\kappa}]=E_{RMF}[{\rho}]+E_{pair}[{\kappa}] \;,
	\end{equation}
	where $ E_{RMF}[{\rho}]$ is the RMF-functional (Relativistic Mean Field). By integrating the Hamiltonian density (\ref{Eq:Ham}) over the $r$-space we obtain
	\begin{equation}
	\label{Eq:EDF-DD}
	E_{RMF}[\rho] = %
	\int d^3r\,{\cal H}(\mathbf{r}).%
	\end{equation}
	and $E_{pair}[{\kappa}]$ is the the pairing part of the RHB functional which is given by:
	\begin{equation}
	\label{Eq:pairing-energy}
	E_{pair}[{\kappa}] = \frac{1}{4}\sum_{n^{}_1n^\prime_1}\sum_{n^{}_2n^\prime_2}
	\kappa^\ast_{n^{}_1n^\prime_1} \langle n^{}_1n^\prime_1|V^{pp}|n^{}_2n^\prime_2\rangle\kappa^{}_{n^{}_2n^\prime_2}.%
	\end{equation}
	with $\langle n^{}_1n^\prime_1|V^{pp}|n^{}_2n^\prime_2\rangle$ are the matrix elements of the two-body pairing interaction. The densities $\rho$ and $\kappa$ are given by eq (\ref{rho}) and eq (\ref{kappa}), successively.
	By the variational principle, we obtain the RHB equation similar to that given by eq (\ref{HFBeq}):
	\begin{eqnarray}
	\left(\begin{array}{cc} h_{D}-\lambda & \Delta \\
	-\Delta^* & -(h_{D}+\lambda)^*
	\end{array} \right)\left(\begin{array}{c} U \\ V \end{array} \right) =E\left(\begin{array}{c} U \\ V \end{array}\right), \, \label{eqdirhb}
	\end{eqnarray}
	Here, $h_{D}$ is the Dirac Hamiltonian for the nucleons with mass m, $\lambda$ is the chemical potential defined by the constraints on the average particle number for protons and neutrons, U and V are quasi-particle Dirac spinors\cite{Kucharek,Afanasjev} and E denotes the quasi-particle energies.
	The Dirac Hamiltonian
	\begin{equation}
	\label{Eq:Dirac0}
	\hat{h}_D = \mathbf{\alpha}(\mathbf{p}-\mathbf{\Sigma}) + \Sigma_0 + \beta (m+\Sigma_s).
	\end{equation}
	contains the attractive scalar potential
	\begin{eqnarray}
		\Sigma_s(\mathbf{r}) &= g_\sigma \sigma(\mathbf{r}),
	\end{eqnarray}
	a repulsive vector potential
	\begin{eqnarray}
		\Sigma_0(\mathbf{r}) &= g_\omega\omega_0(\mathbf{r}) + g_\rho \vec{\tau}\cdot\vec{\rho}_0(\mathbf{r})
		+ eA_0(\mathbf{r}) + \Sigma_0^R(\mathbf{r}).
	\end{eqnarray}
	and a magnetic potential
	\begin{eqnarray}
		\Sigma_\mu(\mathbf{r}) &= g_\omega\omega_\mu(\mathbf{r}) + g_\rho \vec{\tau}\cdot\vec{\rho}_\mu(\mathbf{r})
		+ eA_\mu(\mathbf{r}) + \Sigma_\mu^R(\mathbf{r}).
	\end{eqnarray}

	\section{Details of Calculations}
	This work is realized by employing two methods:  the non relativistic Hartree-Fock-Bogoliubov (HFB) approach based on SLy4 Skyrme functional\cite{Chabanat} by using the HFBTHO (v2.00d)\cite{Stoitsov2013} computer code and the relativistic (covariant) density functional theory (CDFT) based on the DD-ME2 force parametrization\cite{Lalazissis} by using the DIRHBZ\cite{Niksik2014} computer code. 
	\subsection{Method 1: Numerical implementation of the non relativistic HFB equations}
	Presently, the most widely used effective theories are the HFB approach with either Skyrme or Gogny interactions and the relativistic mean field model. We choose the HFB approach with Skyrme force for the present work. In this case, the pairing force, in the particle-particle (pp) channel\cite{Dobaszewski1995,Fayanas,Stoitsov2013}, is given by:
	\begin{equation}	
	v_{pair}^{n,p}=V_0^{n,p}\left[1-\alpha\left(\frac{\rho(r)}{\rho_c}\right)^{\beta}\right]\delta(r-r')
	\end{equation}
	where $\rho(r)$ is the local density and $\rho_c=0.16fm^{-3}$ is the saturation density of symmetric infinite nuclear matter (INM). The factor $\alpha$  enables one to change the properties of the pairing force: if it is equal to 0, the pairing force has pure volume character and does not depend on the isoscalar density; if it is set to 1, the pairing force is only active at the surface; if is fixed to 0.5, the pairing force has mixed volume-surface characteristics. Here we choice $\alpha$=0.5.
	The parameter $V_0^{n,p}$ is the value of the pairing strength for protons and neutrons which can be adjusted phenomenologically by fitting. In order to avoid nonphysical divergences, the definition of the force involves also an energy cut-off parameter in the valence single-particle space to limit the active pairing space above the Fermi level to one major shell. In the present study we have used the energy cut-off parameter equal to 60~MeV, and we have employed the SLy4 parametrization \cite{Chabanat}. This parametrization has been introduced by the Saclay-Lyon collaboration in the 90's. It performs well for the total energies, radii, and moments, and it is also reliable	when it comes to predictions of long isotopic sequences. Table \ref{table 1} summarizes the parameters of this parametrization.\\
%%%%%%%%%%%%%%%%%%%%%%%%%%%%%%%%%%%%%%%%%%%%%%%%%%%%%%%%%%%%%%%%%%%%%%%%
	\begin{table}[ht]
		\caption{Parameters of the skyrme force (SLy4).}\label{table 1}
			\begin{center}
	{\begin{tabular}[c]{@{}c@{\hspace{18pt}}@{\hspace{18pt}}c@{}} 
				\hline
				Parameter & SLy4 parametrization \\ 
				\hline
				t$_0$ (MeV fm$^3$)   &     -2488.91 \\
				t$_1$ (MeV fm$^5$)   &       486.82 \\
				t$_2$ (MeV fm$^5$)   &      -546.39 \\
				t$_3$ (MeV fm$^4$)   &       13777.0 \\
				x$_0$ 		    &       0.834 \\
				x$_1$    	    &       -0.344 \\
				x$_2$   	    &      -1.0 \\
				x$_3$  		    &       1.354 \\
				W$_0$ (MeV fm$^3$)   &       123	\\
				$\sigma$            &        1/6    \\  
				\hline
			\end{tabular}}
				\end{center}
		\end{table}
%%%%%%%%%%%%%%%%%%%%%%%%%%%%%%%%%%%%%%%%%%%%%%%%%%%%%%%%%%%%%%%%%%%%%%%%
		We recall that the aim of our work is to investigate the magicity in neutron-rich argon and calcium isotopes. For this purpose we have used the code HFBTHO v2.00d\cite{Stoitsov2013} that solves the Skyrme Hartree-Fock  (HF) or Skyrme Hartree-Fock-Bogoliubov (HFB) equations by using the cylindrical transformed deformed harmonic oscillator basis. This program iteratively diagonalizes the Hartree-Fock-Bogolyubov Hamiltonian based on generalized Skyrme-like energy densities and zero-range pairing interactions until a self-consistent solution is found.
				
	To study the convergence of the HFBTHO results in nuclei under investigation, we calculated the binding energy $BE$ and the neutron rms radii $r_n$ as functions of the number of shells $N_{sh}$ for the neutron-rich nuclei $^{60}Ca$ and $^{50}Ar$ (see Figure~\ref{Nsh}). 
		%%%%%%%%%%%%%%%%%%%%%%%%%%%%%%%%%%%%%%%%%%%%%%%%%%%%%%%%%%%%%%%%%%%%%%%%
		\begin{figure}[h!]
			\begin{minipage}{0.45\textwidth}
				\centering \includegraphics[scale=0.5]{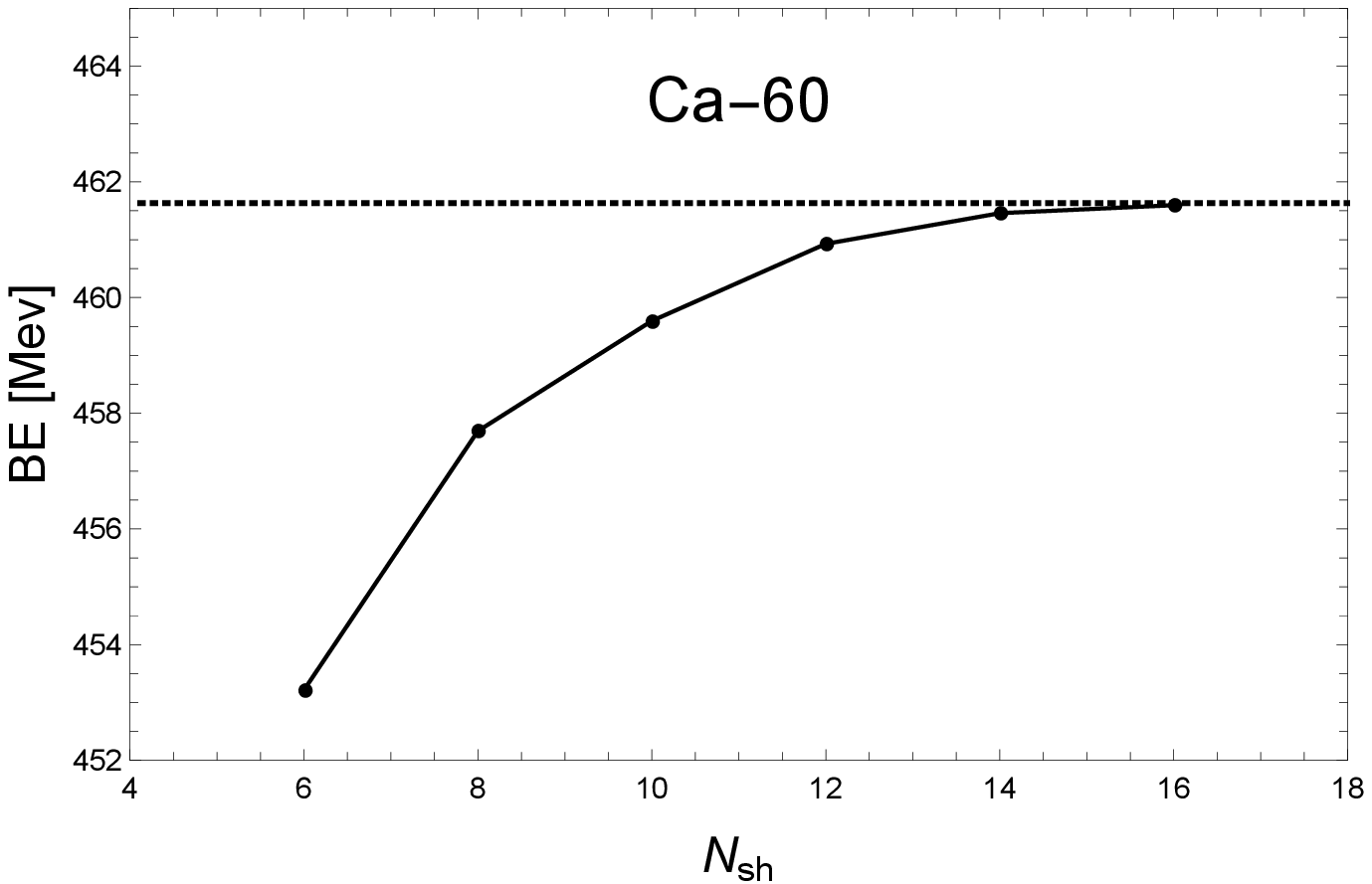}
				%  \caption{}
			\end{minipage}\hfill
			\begin{minipage}{0.45\textwidth}
				\centering \includegraphics[scale=0.5]{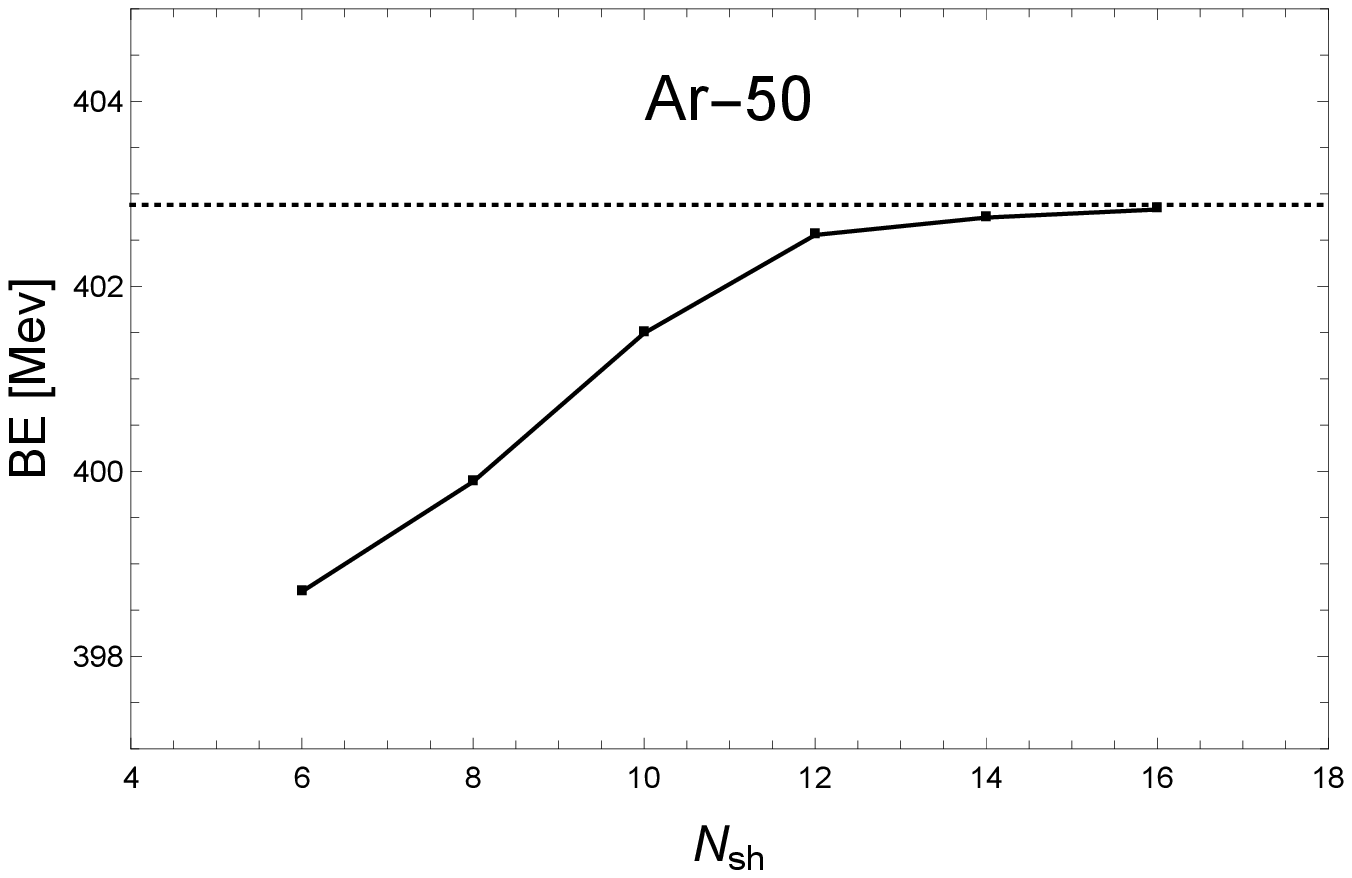}
				%  \caption{}
			\end{minipage}\hfill
				\begin{minipage}{0.45\textwidth}
					\centering \includegraphics[scale=0.5]{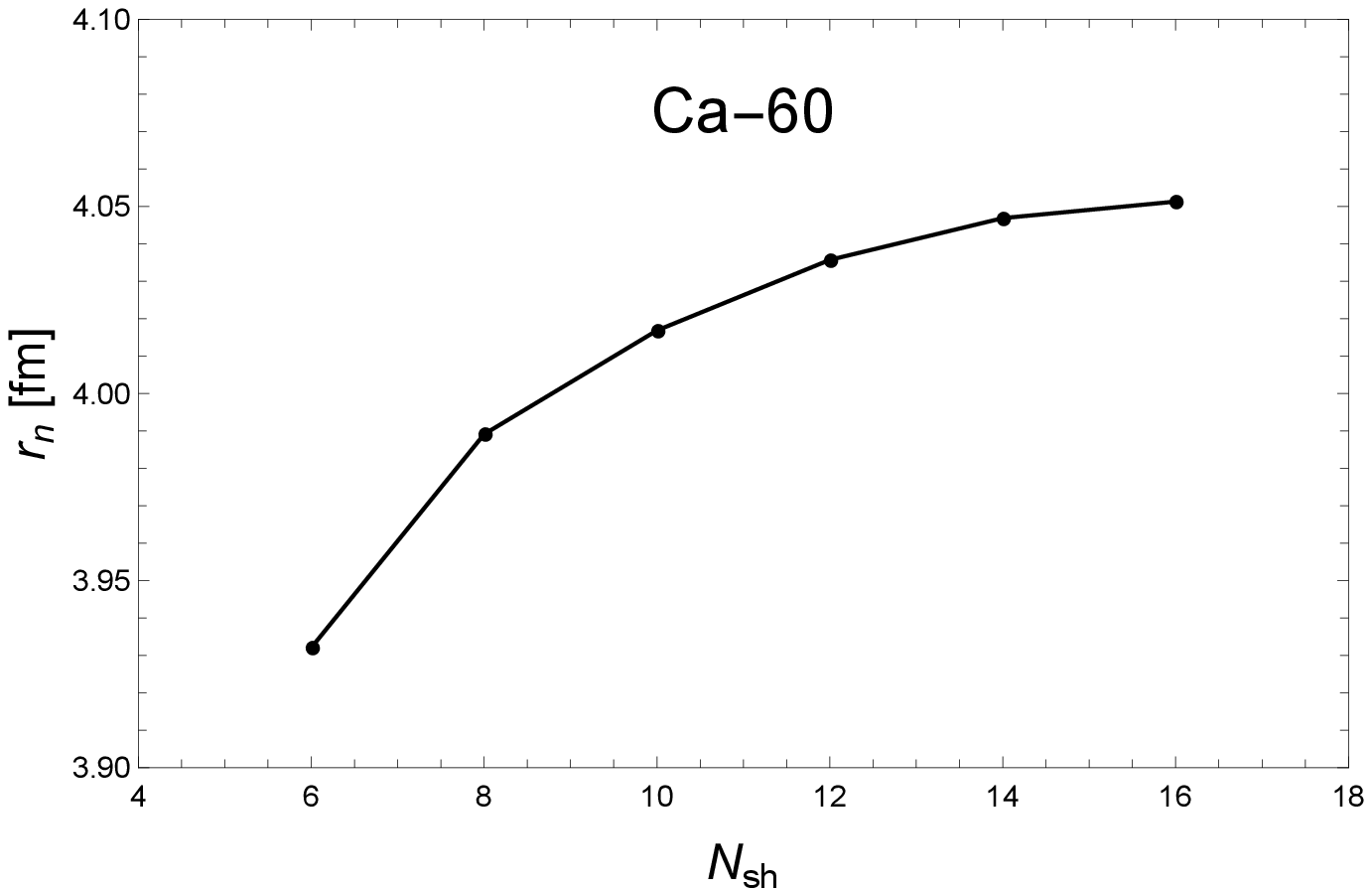}
					%  \caption{}
				\end{minipage}\hfill
				\begin{minipage}{0.45\textwidth}
					\centering \includegraphics[scale=0.5]{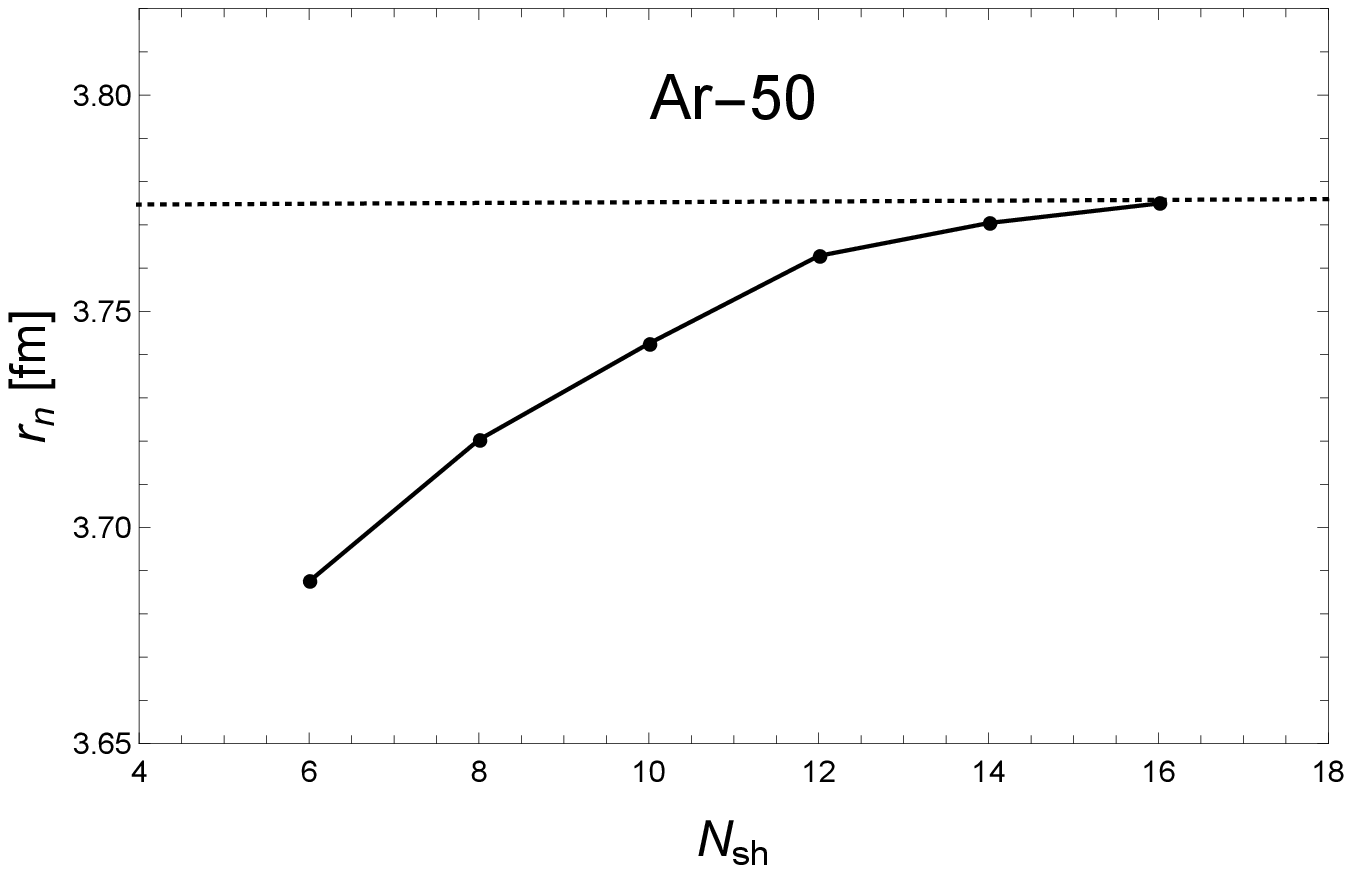}
					%  \caption{}
				\end{minipage}\hfill
			\caption{Binding energies $BE$ (top panels) and the neutron rms radii (bottom panels) in the HFB+SLy4 calculations for $^{60}Ca$ (left panels) and $^{50}Ar$ (right panels) as functions of the number of shells $N_{sh}$.\label{Nsh}}
		\end{figure}
		%%%%%%%%%%%%%%%%%%%%%%%%%%%%%%%%%%%%%%%%%%%%%%%%%%%%%%%%%%%%%%%%%%%%%%%%
        As shown in this Figure, when we increase $N_{sh}$, both the binding energy and the neutron rms radii converge at $N_{sh}=14$. Therefore, all performed calculations by using the code HFBTHO v2.00d have been carried out in a full spherical basis of $N_{sh}=14$ oscillator shells (the number of basis states is $N_{states} = 680$). The oscillator frequency was fixed at $1.2 \hbar \omega$ for $\hbar \omega_0= 41/A^{1/3} MeV$.
    
       The HO basis implies that there are no continuum states modeled. And as the convergence of the results is confirmed, the effect of the continuum state is only a minor issue for most of the present results, but it could affect the results for the drip-line Ar isotopes near N=40.  Many authors have studied the role of the coupling to the continuum, see, e.g., the continuum-HFB results reported in ref. \cite{Xia} and the review of Okolowicz et al. in ref. \cite{Okolowicz} in the context of the shell model, or the review by Forssen et al in ref \cite{Forssen}.\\
		The value of the deformation $\beta$ is taken from the column $\beta_2$ of the Ref \cite{Moller}. The number of Gauss-Laguerre and Gauss-Hermite quadrature points was $N_{GL} = N_{GH} = 40$, and the number of Gauss-Legendre points for the integration of the Coulomb potential was $N_{Leg} = 80$.\\
		In this work the pairing strength $V_0^{p,n}$ for both protons and neutrons has been adjusted to reproduce well the experimental binding energies, then by fitting the obtained values of $V_0^{p,n}$ to N, we have found the following formulas appropriate for each one among the nuclei under investigation:
		\begin{equation}
		V_0^{p,n}=\cases{270 & ~ for Ar isotops\\
		5N-30 & ~ for Ca isotops}
			\end{equation}
		where N is the neutron number. For more details, see Ref \cite{Elbassem} and references therein.
		
		\subsection{Method 2: Numerical  implementation of the RHB equations}
		Here, the used computer code  is  DIRHBZ\cite{Niksik2014}. Analogously  to  HFBTHO code\cite{Stoitsov2013}, the RHB equation is solved in the configurational space of harmonic oscillator wave functions with appropriate symmetry, whereas the densities are computed in coordinate space.\\
		By the same principle previously used in the case of HFTHO, we found that the DIRHBZ results converge when the numbers  of oscillator shells for fermions and bosons are set to $N_F=12$ and $N_B=20$, respectively (see Figure~\ref{Nshcdft}). The $\beta$-deformation parameter for the harmonic oscillator basis as well as for the initial Woods-Saxon potential is set to 0.
		The method can be applied to spherical, axially and non-axially deformed nuclei. Here, we show  results of calculations for ground states properties using the axially symmetric quadrupole deformation based on the effective interaction density-dependent meson-exchange (DD-ME2). The parameters of this interaction are given in table~\ref{table 2}:
%%%%%%%%%%%%%%%%%%%%%%%%%%%%%%%%%%%%%%%%%%%%%%%%%%%%%%%%%%%%%%%%%%%%%%%%
		\begin{table}[h]
		\caption{The parameters DD-ME2 interaction. The masses are given in MeV and all other parameters are dimensionless.\label{table 2}}
		\begin{center}
				{\begin{tabular}{@{}c@{\hspace{18pt}}@{\hspace{18pt}}c@{}} \hline
					Parameter      & DD-ME2 interaction \\ \hline
					m                   &    939 \\
					$m_{\sigma}$  &       550.124 \\
					$m_{\omega}$ &      783.000 \\
					$m_{\rho}$      &      763.00 \\
					$m_{\delta}$      &      0.000 \\
					$g_{\sigma}$   &       10.5396 \\
					$g_{\omega}$   &       13.0189 \\
					$g_{\rho}$        &       3.6836 \\
					$g_{\delta}$        &       0.000 \\
					$a_{\sigma}$    &       1.3881 \\
					$b_{\sigma}$    &       1.0943 \\
					$c_{\sigma}$    &       1.7057 \\
					$d_{\sigma}$    &       0.4421 \\
					$e_{\sigma}$    &       0.4421 \\
					$a_{\omega}$   &       1.3892 \\
					$b_{\omega}$   &       0.9240 \\
					$c_{\omega}$   &       1.4620 \\
					$d_{\omega}$   &       0.4775 \\
					$e_{\omega}$   &       0.4775 \\
					$a_{\rho}$        &       0.5647 \\  \hline
				\end{tabular}}
			\end{center}
			\end{table}
%%%%%%%%%%%%%%%%%%%%%%%%%%%%%%%%%%%%%%%%%%%%%%%%%%%%%%%%%%%%%%%%%%%%%%%%
%%%%%%%%%%%%%%%%%%%%%%%%%%%%%%%%%%%%%%%%%%%%%%%%%%%%%%%%%%%%%%%%%%%%%%%%
		\begin{figure}[h!]
			\begin{minipage}{0.45\textwidth}
					\centering \includegraphics[scale=0.5]{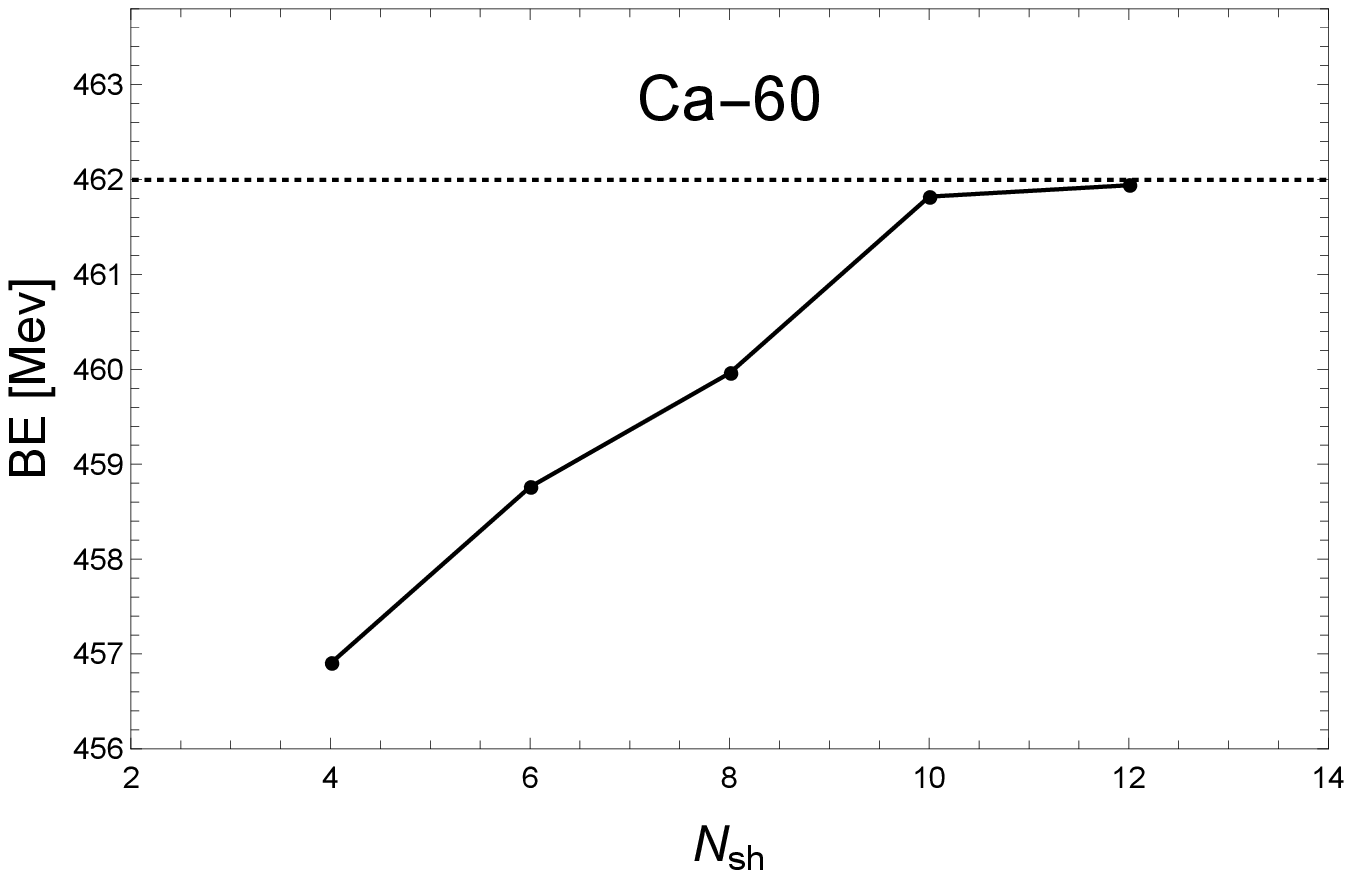}
						%  \caption{}
			\end{minipage}\hfill
			\begin{minipage}{0.45\textwidth}
					\centering \includegraphics[scale=0.5]{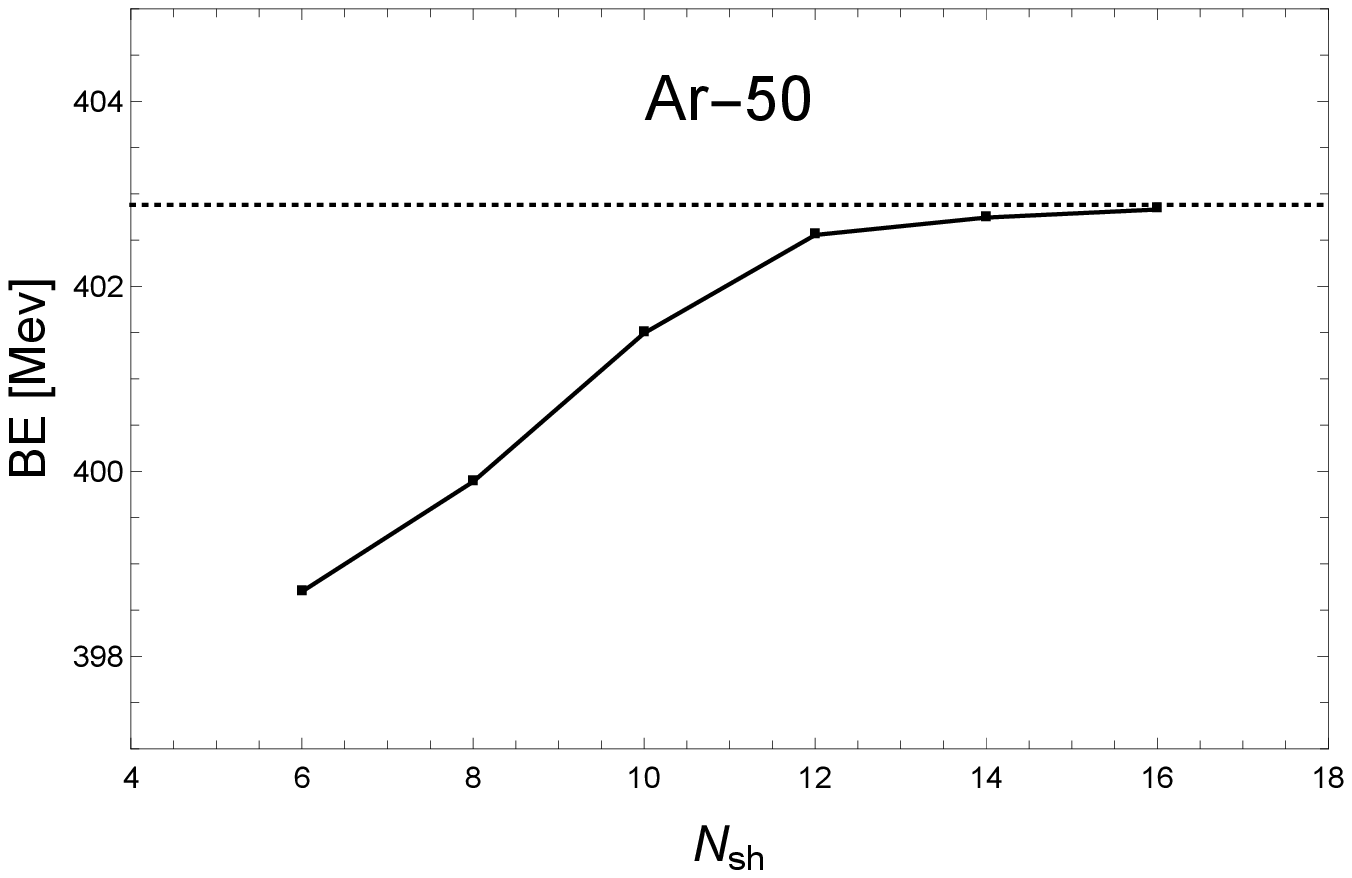}
						%  \caption{}
			\end{minipage}\hfill
			\begin{minipage}{0.45\textwidth}
					\centering \includegraphics[scale=0.5]{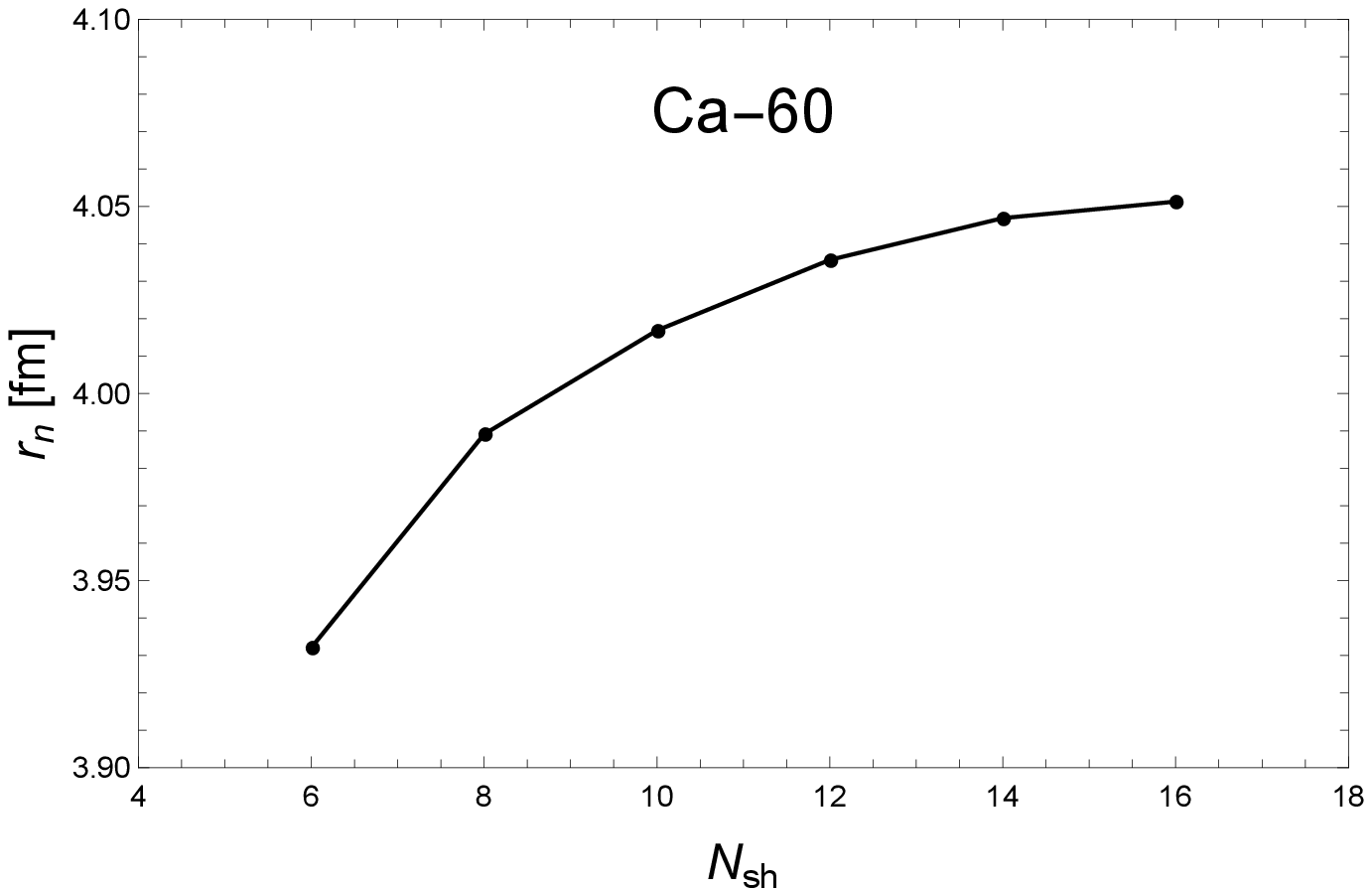}
						%  \caption{}
			\end{minipage}\hfill
			\begin{minipage}{0.45\textwidth}
					\centering \includegraphics[scale=0.5]{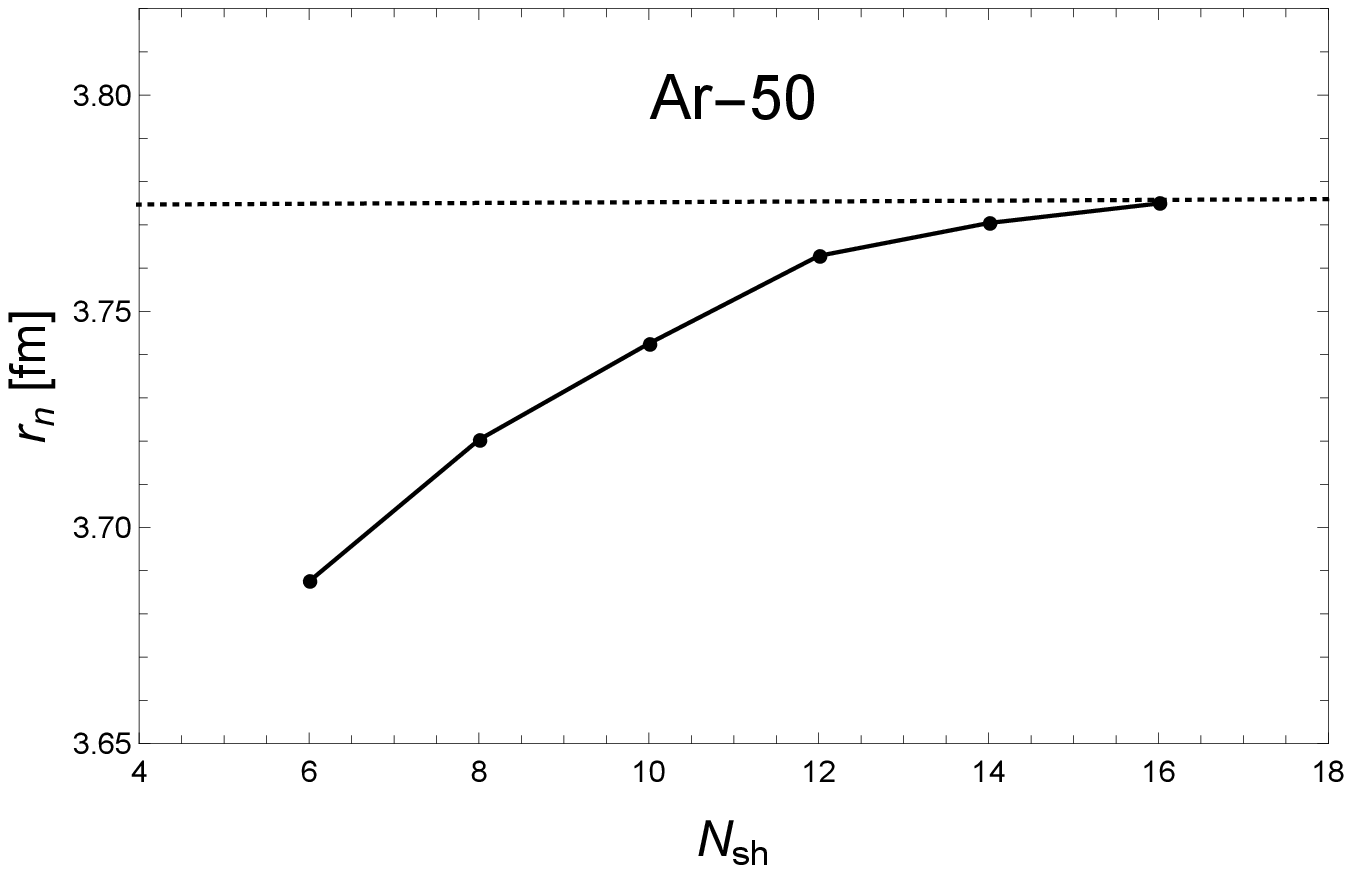}
						%  \caption{}
			\end{minipage}\hfill
	\caption{Binding energies $BE$ (top panels) and the neutron rms radii (bottom panels) in the CDFT+DDME2 calculations for $^{60}Ca$ (left panels) and $^{50}Ar$ (right panels) as functions of the number of shells $N_{sh}$.\label{Nshcdft}}
			\end{figure}
					%%%%%%%%%%%%%%%%%%%%%%%%%%%%%%%%%%%%%%%%%%%%%%%%%%%%%%%%%%%%%%%%%%%%%%%%
			\section{Results and Discussion}
			Recent measurements   of exotic nuclei features at facilities for radioactive ion beams have revealed that the magic numbers may change locally in those exotic nuclei leading to  disappearance of classic magic numbers (2, 8, 20, 28, 50 and 126)\cite{Goeppert} and  appearance of new ones.
			
			Several quantities allow  describing well the shell closure in neutron rich nuclei,  among which we can cite: one and two-neutron separation energies $S_{n}$ and $S_{2n}$, two-neutron shell gap $\delta_{2n}$, neutron pairing gap $\Delta_{3n}$ and pairing energy. The existence of a magic number is signed either by the sudden jump in $S_{2n}$ or by a spectacular increase of $\delta_{2n}$ and of $\Delta_{3n}$. Moreover, the pairing energy vanishes for magic nuclei.
			
			In our work we used, for comparison, the Finite Range Droplet Model (FRDM)\cite{FRDM} and the available experimental data\cite{Exp}.
			
			\subsection{Two-neutron separation energies }\label{StwoN}
			The first indication for shell closure is two-neutron separation energy $S_{2n}$ which is defined by:
			\begin{equation}
			S_{2n}(Z,N)=B(Z,N)-B(Z,N-2)
			\end{equation}
			where B(Z,N) is the positive value of binding energy of a nucleus with Z protons and N neutrons.
			Figure~\ref{S2n} displays the variation of two-neutron separation energy for argon and calcium isotopes as a function of the neutron number N obtained in HFB calculations where SLy4 parametrization has been used and in CDFT calculations based on DD-ME2 force, and we compare them with the FRDM model and the available experimental data taken from Ref \cite{Exp}. From this figure, the HFB+SLy4 and the CDFT+DD-ME2 calculations and FRDM predictions give, approximately, similar results in agreement with the experimental data. Except for Ar, we observe that the microscopic calculations predict N=20 as closure, but the FRDM and the experiment suggest that N=18 is a magic number, as clearly seen on the Figure~\ref{S2n}. Also, it is clear that two separation energy $S_{2n}$ presents a remarkable jump at the known classic magic number N=28 for $_{20}Ca$ and $_{18}Ar$ isotopes. A similar behavior is observed  around N = 32 and N=40 for both investigated nuclei. This behavior corresponds to the appearance of closed shells around N=32 and N=40.
			
			Furthermore, the neutron number N=14 observed in oxygen (see Ref \cite{Janssens}) as a magic number, is also confirmed by our calculations. In contrast, in calcium and argon isotopes no shell effects can be seen at N = 16 and 26, that were predicted as a shell gap in oxygen\cite{Yu}.
			
			It is important to note that FRDM predictions fail to reproduce some shell closures such as N=14, N=32 and N=40.
			
%%%%%%%%%%%%%%%%%%%%%%%%%%%%%%%%%%%%%%%%%%%%%%%%%%%%%%%%%%%%%%%%%%%%%%%%
			\begin{figure}[h!]
				\begin{minipage}{0.45\textwidth}
					\centering \includegraphics[scale=0.5]{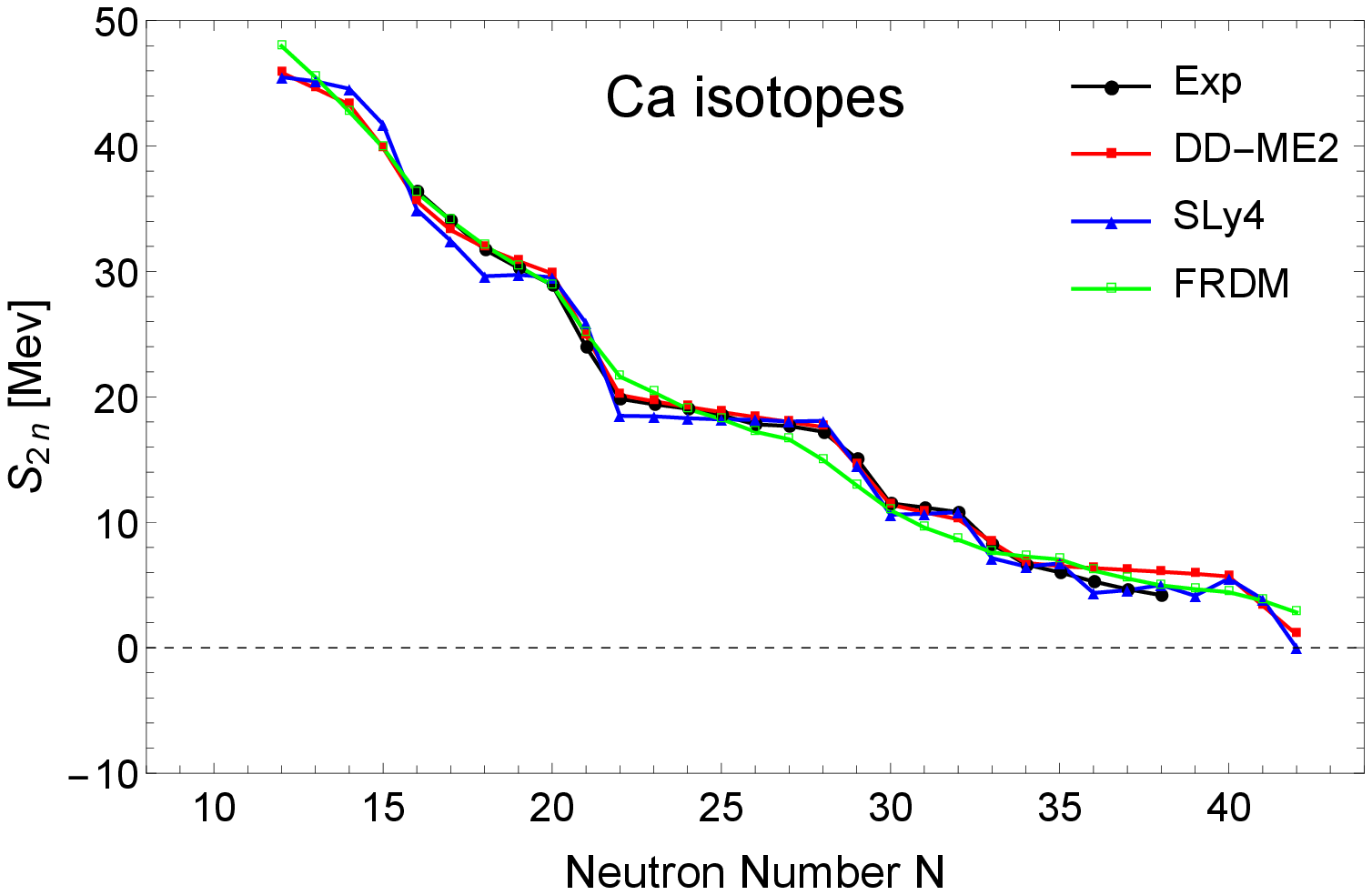}
					%  \caption{}
				\end{minipage}\hfill
				\begin{minipage}{0.45\textwidth}
					\centering \includegraphics[scale=0.5]{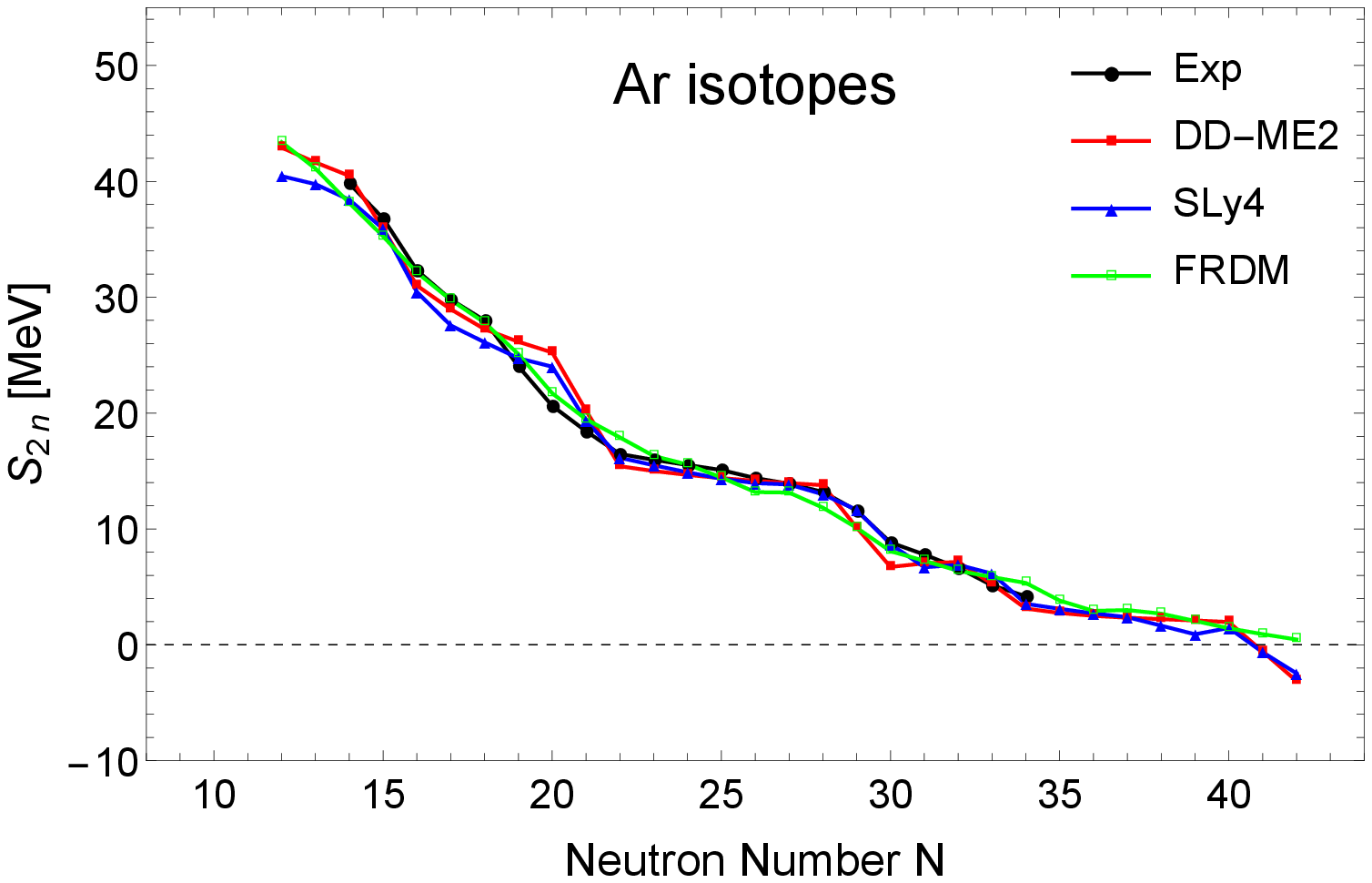}
					%  \caption{}
				\end{minipage}\hfill
				\caption{Two-neutron separation energies as a function of neutron number N for  Ca isotopes (left panel) and Ar isotopes (right panel), obtained by non relativistic HFB with SLy4 parametrization and CDFT with DD-ME2 force calculations, and compared with FRDM predictions and available experimental data.\label{S2n}}
			\end{figure}
%%%%%%%%%%%%%%%%%%%%%%%%%%%%%%%%%%%%%%%%%%%%%%%%%%%%%%%%%%%%%%%%%%%%%%%%
			\subsection{One-neutron separation energy}\label{SoneN}
			Another physical quantity that can show discontinuities (drop) when crossing a shell closure is the one-neutron separation energy $S_{n}$, defined as:
			\begin{equation}
			S_{n}(Z,N)=B(Z,N)-B(Z,N-1)
			\end{equation}
			Figure~\ref{fig:Snf} shows one-neutron separation energies $S_{n}$ as a function of the neutron number. Our HFB and CDFT results are compared with predictions of FRDM model and experimental data. In general, the agreement between our calculations and  FRDM theory and available experimental data is nicely good.
			
			For the results shown in Figure~\ref{fig:Snf}, one can observe a significant drop in the separation energies for the well known classic magic numbers N=20 and N=28 in $_{20}Ca$ and $_{18}Ar$ isotopes. Also, one can see a clearly jump in $S_{n}$ values at N=32 and 40 in these nuclei.
			
			These observations indicate a change of structure in exotic region and confirm the appearance of new closed shells at N=32 and N=40. 
			
			Again, the closed shell around N=14 is also observed, while for N=16 and 26 is not observed for both isotopic chains.
			
			Note that local shifts in separation energies can sometimes also be the result of other causes, e.g. the competition and mixing of different nuclear shapes at low energies.
			Thus, this quantity alone is not always sufficient to establish  new closed shells. However, important additional information can be gained from the study of pairing effect.
%%%%%%%%%%%%%%%%%%%%%%%%%%%%%%%%%%%%%%%%%%%%%%%%%%%%%%%%%%%%%%%%%%%%%%%	
			\begin{figure}[h!]
				\begin{minipage}[b]{0.45\linewidth}
					\centering \includegraphics[scale=0.5]{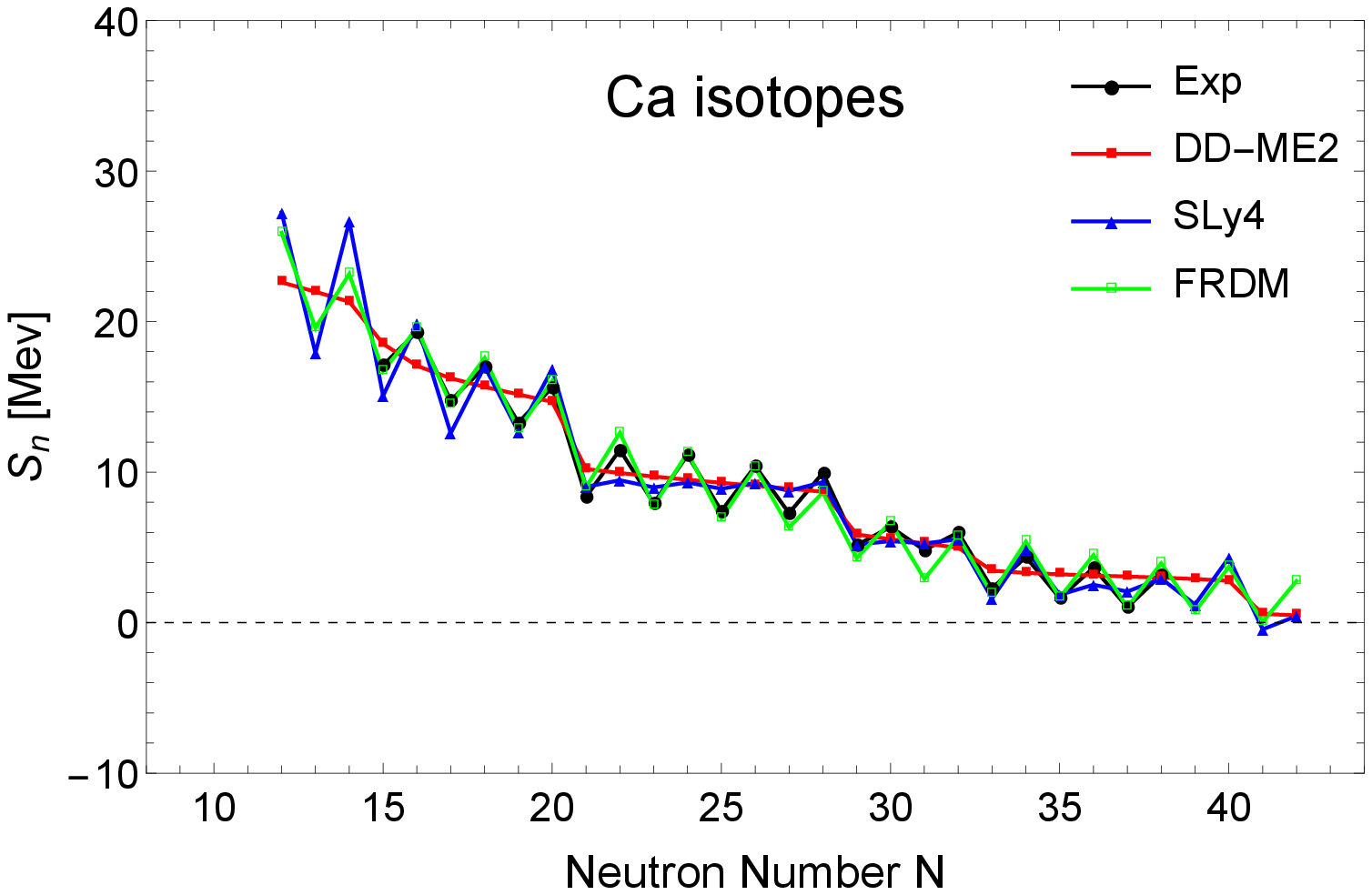}
					%  \caption{}
				\end{minipage}\hfill
				\begin{minipage}[b]{0.45\linewidth}
					\centering \includegraphics[scale=0.5]{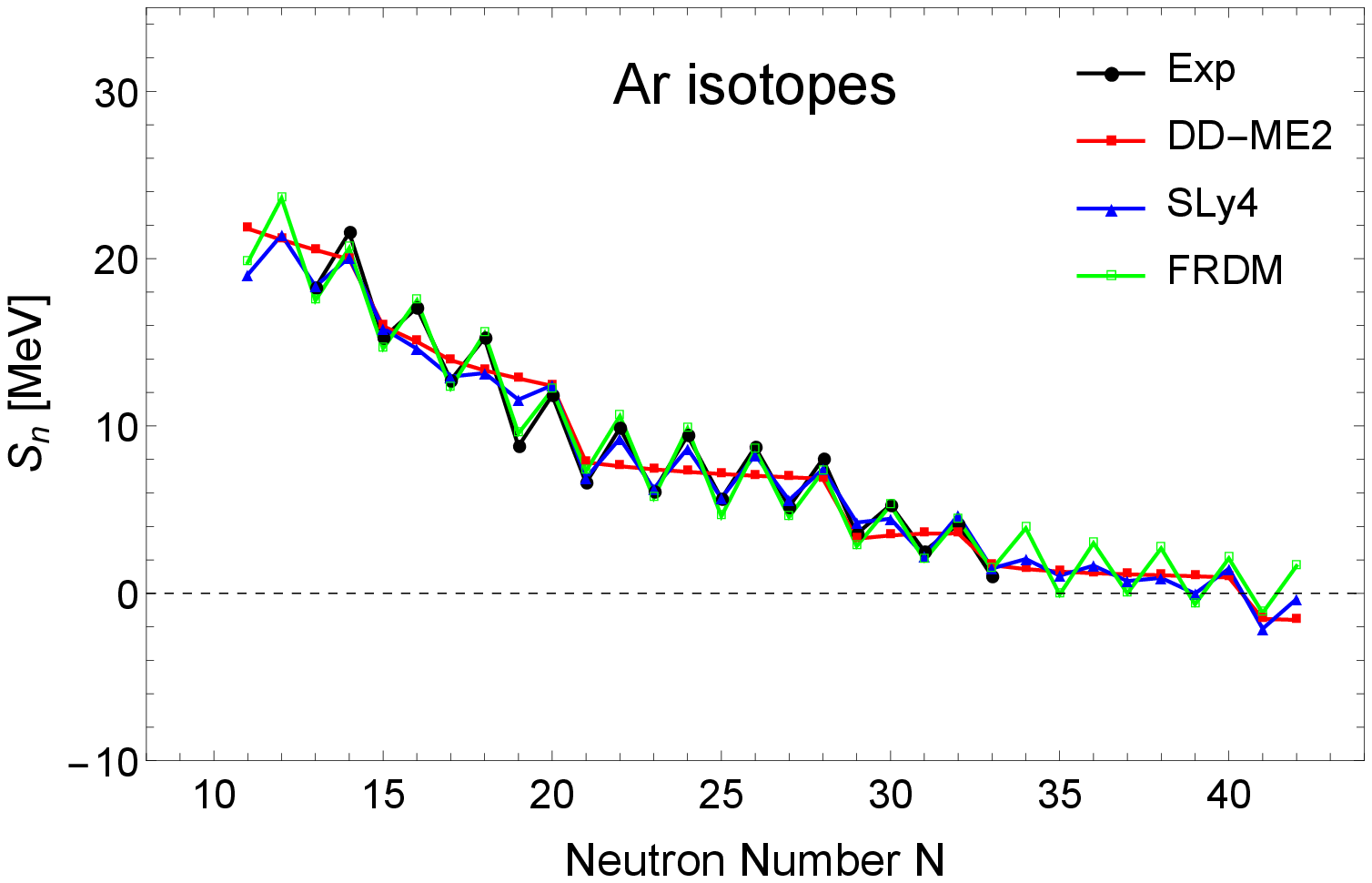}
					%  \caption{}
				\end{minipage}\hfill
				\caption{One-neutron separation energy as a function of neutron number N for  Ca isotopes (left panel) and Ar isotopes (right panel), obtained by non relativistic HFB with SLy4 parametrization and CDFT with DD-ME2 force calculations, and compared with FRDM predictions and available experimental data}\label{fig:Snf}
			\end{figure}
%%%%%%%%%%%%%%%%%%%%%%%%%%%%%%%%%%%%%%%%%%%%%%%%%%%%%%%%%%%%%%%%%%%%%%%%
			\subsection{Two-neutron shell gap}\label{Shellgap}
			A more direct measure of a shell closure is the observation of a peak in the two-neutron shell gap $\delta_{2n}$:
			\begin{eqnarray}
				\delta_{2n}&=2B(Z,N)-B(Z,N-2)-B(Z,N+2)\\ \nonumber
				&=S_{2n}(Z,N)-S_{2n}(Z,N+2)
			\end{eqnarray}
			where B and $S_{2n}$ are the binding energy and the two-neutron separation energy, respectively.
			
			In the left panel of Figure~\ref{D2}, we show the $_{20}Ca$ shell gap $\delta_{2n}$ as a function of N. The maximum values at the classic magic numbers N=20 and N=28 are clearly visible. From this figure one can see also a sharp increase in $\delta_{2n}$ at neutron numbers  N=14, N=32 and N=40 corresponding to the appearance of new magic numbers in calcium.
			
		The right panel of the Figure~\ref{D2} shows shell gap as a function of N for argon nucleus. In this case, one can clearly see the peak in $\delta_{2n}$, obtained by all calculations, for the well known classic magic number N=28 and for the neutron numbers N=14, N=32 and N=40 which make them as new magic numbers. However, around N=16 and 26, no closed shells are observed.							
		
		Note that the peak at N=18 observed in the Ar data (Figure~\ref{D2},right panel) is not reproduced by the microscopic models.
% % % % % % % % % % % % % % % % % % % % % % % % % % % % % % % % 	
			\begin{figure}[h!]
				\begin{minipage}[b]{0.45\linewidth}
					\centering \includegraphics[scale=0.5]{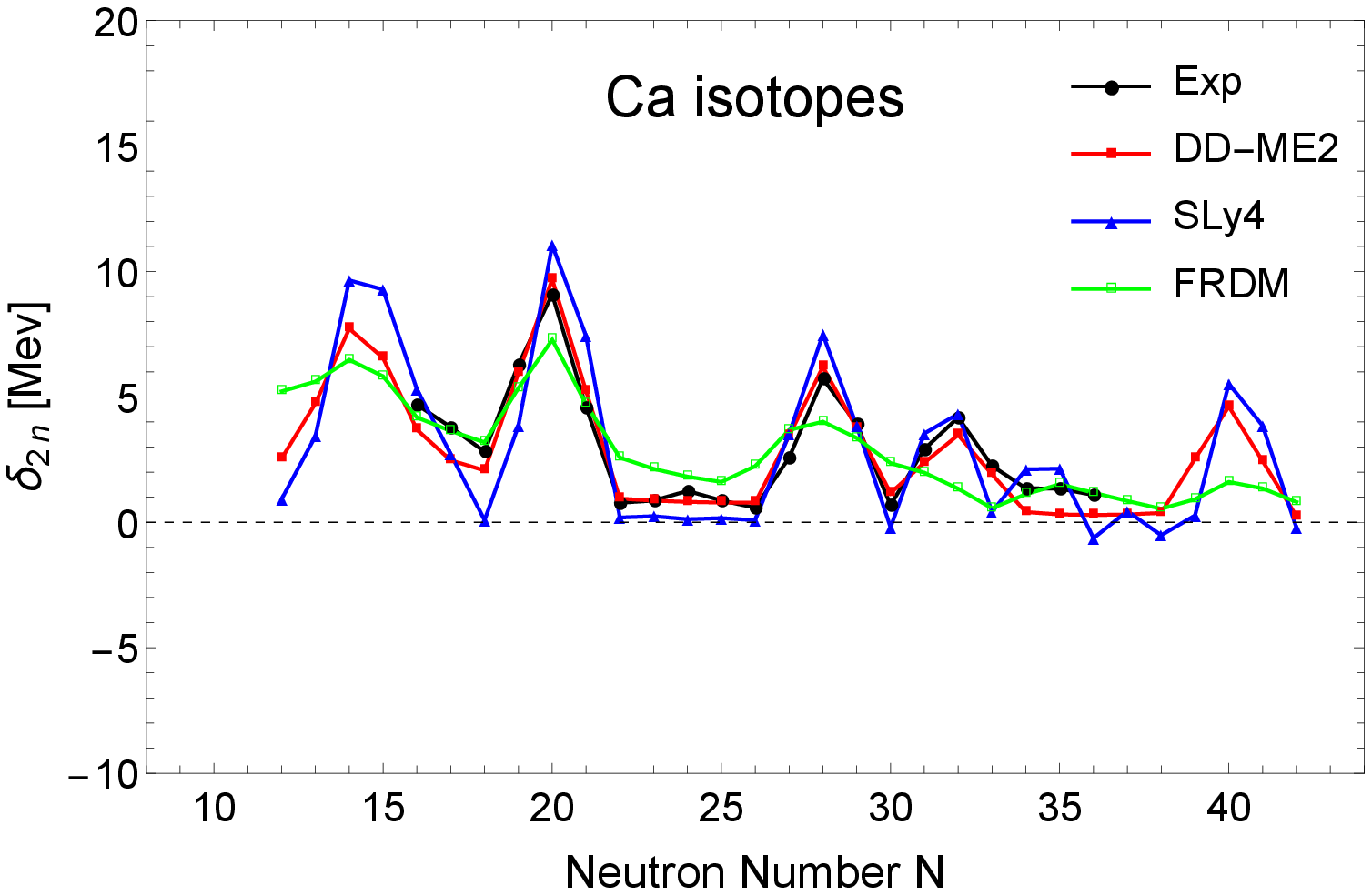}
					%  \caption{}
				\end{minipage}\hfill
				\begin{minipage}[b]{0.45\linewidth}
					\centering \includegraphics[scale=0.5]{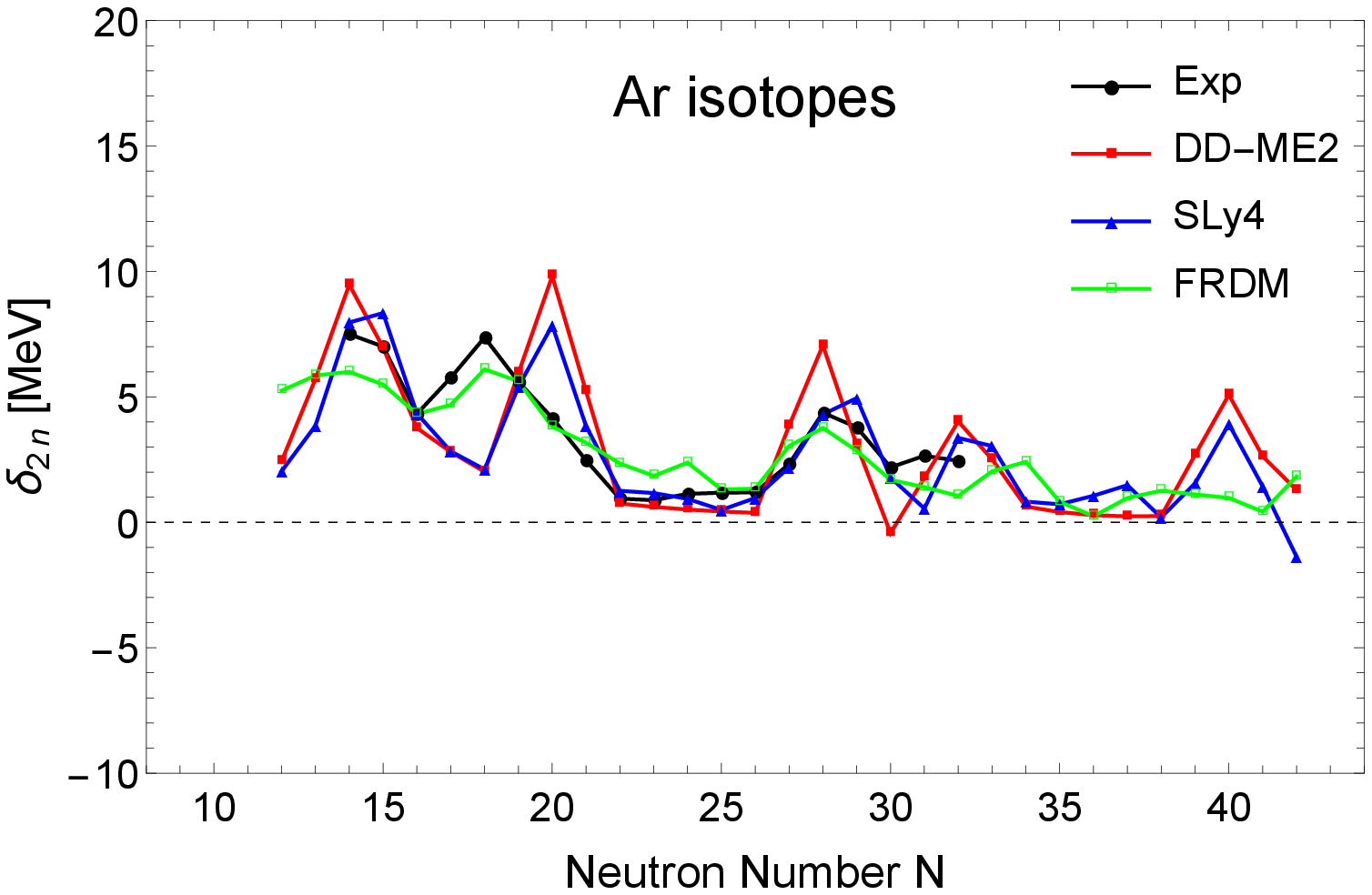}
					%  \caption{}
				\end{minipage}\hfill
				\caption{Calcium (left panel) and argon (right panel) two-neutron shell gap calculated by using HFB+SLy4 and CDFT+DD-ME2, and compared with experimental data and FRDM predictions.}\label{D2}
			\end{figure}
% % % % % % % % % % % % % % % % % % % % % % % % % % % % % % % % %	
			\subsection{Neutron pairing gap}\label{Pairinggap}
			The stability of nuclei with a magic number of protons or neutrons has direct consequences on the pairing gap. This quantity is defined by an approximate formula so-called the three-point gaps $\Delta_{3n}$\cite{Bender}:
			\begin{eqnarray}\label{Eq3gap}
				\Delta_{3n}&=\pi_N\Bigg[B(Z,N)-\frac{B(Z,N-1)+B(Z,N+1)}{2}\Bigg]\\ \nonumber
				&=\frac{\pi_N}{2}[Sn(Z,N)-Sn(Z,N+1)]
			\end{eqnarray}
			where $\pi_N=(-1)^N$ is the parity number, B and $S_n$ are the binding energy and the one-neutron separation energy, respectively. Note that for even-even nuclei, the  shell closure implies a large pairing gap.
			In Figure~\ref{D3}, the three-point gaps obtained in our calculations, HFB+SLy4 and CDFT+DD-ME2, are compared to  predictions of FRDM and the experimental values, which are calculated from the atomic mass evaluation taken from Ref \cite{Exp} by using the eq (\ref{Eq3gap}). A good agreement between theory and experiment can be clearly seen for both argon and calcium isotopes
			
			In Figure~\ref{D3}, an abrupt increase can be clearly seen at N=14, N=28, N=32 and N=40 for both nuclei under investigation.
			From these results, it follows that the conventional shell closure N=28 persists in $_{20}Ca$ and $_{18}Ar$. Furthermore, the neutron numbers N=32 and N=40, in these nuclei, are  new magic numbers that appear in the exotic region. Also, as discussed in section \ref{StwoN}, the magic number N=14 was reproduced, and N=16 and 26 are not observed. 
			
		 Note that, from Figures 2, 3 and 4 (right panel): Experimentally, N = 20 is not a magic number in the case of argon isotopes. This result has been reproduced by the FRDM model. This can be explained by the fact that the inversion of the standard sd-shell configuration and pf-shell intruder configuration as it has been proved in Refs. \cite{Kanungo,Chaudhuri} in the case of $^{32}Mg$. On the other hand, the microscopic theories show that N = 20 is a magic number. It seems rather that these theories  are failing in this respect. This observation paves the way for a future investigation.
%%%%%%%%%%%%%%%%%%%%%%%%%%%%%%%%%%%%%%%%%%%%%%%%%%%%%%%%%%%%%%%%%%%%%%%%%%%%%
			\begin{figure}[h!]
				\begin{minipage}[b]{0.45\linewidth}
					\centering \includegraphics[scale=0.5]{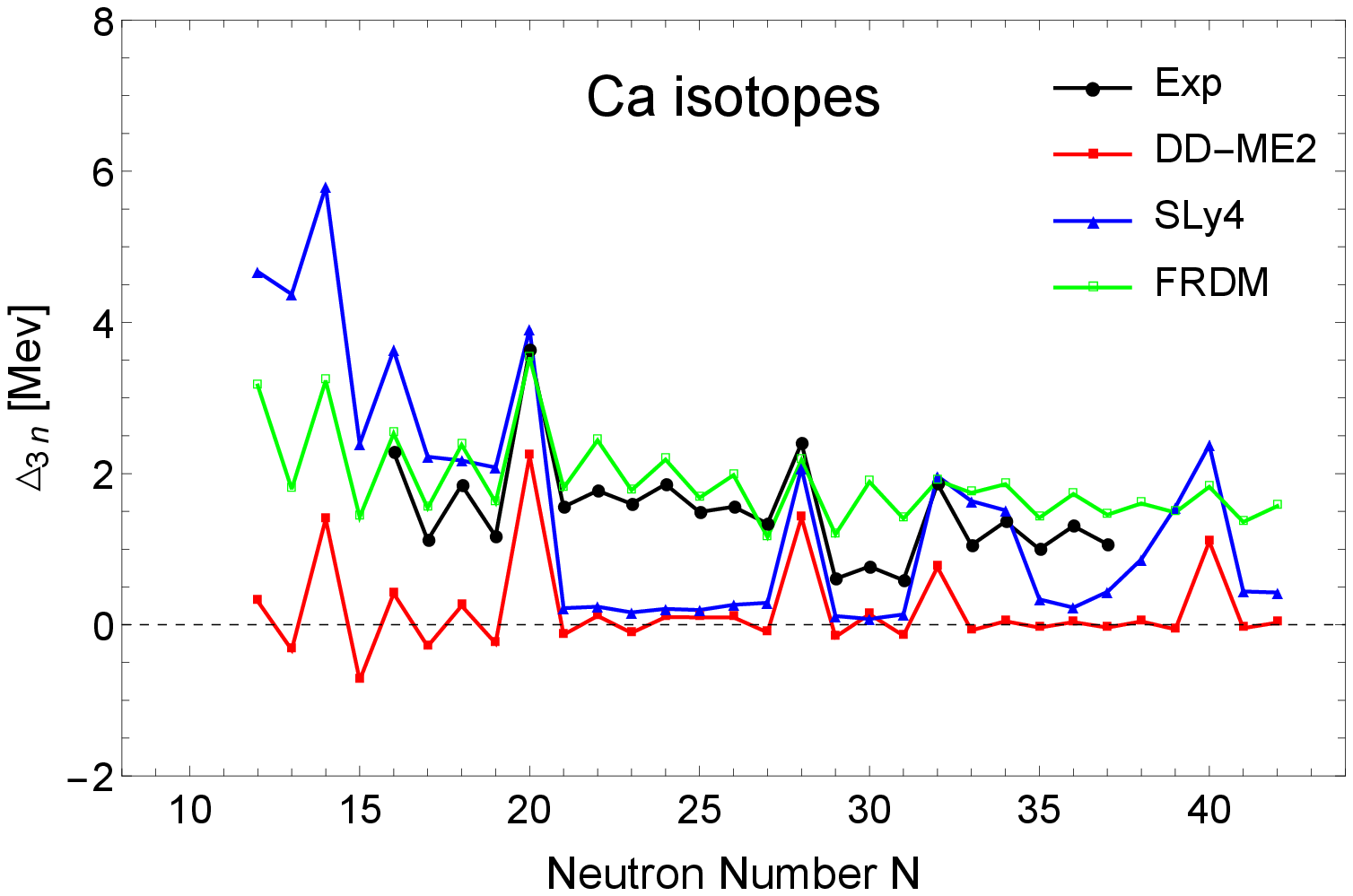}
					%  \caption{}
				\end{minipage}\hfill
				\begin{minipage}[b]{0.45\linewidth}
					\centering \includegraphics[scale=0.5]{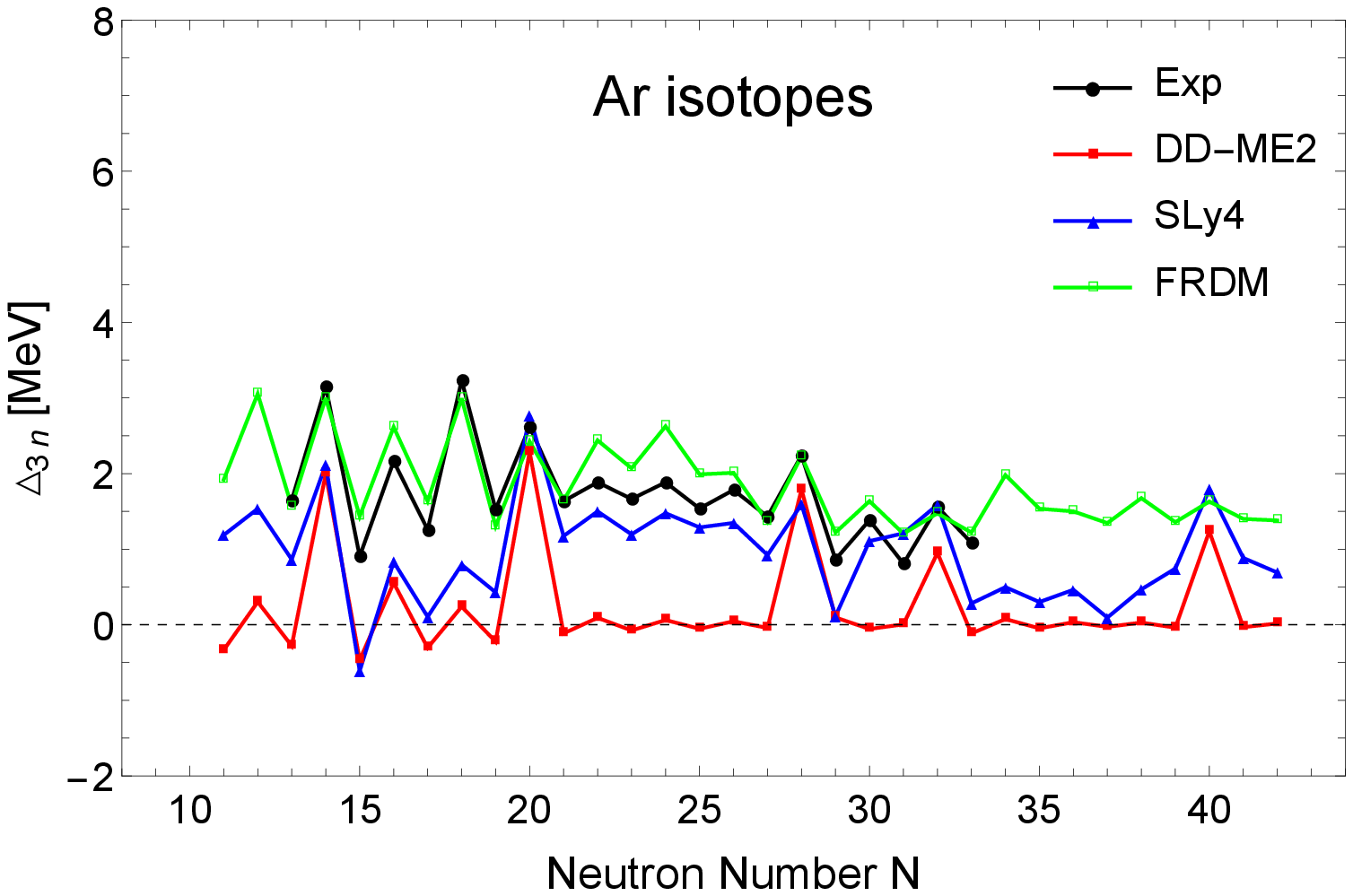}
					%  \caption{}
				\end{minipage}\hfill
				\caption{Values of neutron pairing gap, in MeV, for the studied nuclei .}\label{D3}
			\end{figure}
%%%%%%%%%%%%%%%%%%%%%%%%%%%%%%%%%%%%%%%%%%%%%%%%%%%%%%%%%%%%%%%%%%%%%%%	
			\subsection{Neutron pairing energy}\label{Pairing energy}
			Another method to test the magicity of nuclei is the calculation of the pairing energy. This energy is given by
			\begin{equation}
			E_{pair}=-\frac{1}{2}Tr(\Delta\kappa)
			\end{equation}
			with $\kappa$ is the pairing tensor and $\Delta$ is the pairing field\cite{Stoitsov}. 
			This energy is very low for magic nuclei and reaches high values for deformed ones.
			
			In Figure~\ref{Epair}, we show the neutron pairing energy for calcium (left panel) and for argon (right panel) as a function of N. The two calculations, HFB+SLy4 and CDFT+DD-ME2, produce very similar results.
			
			As shown in Figure~\ref{Epair}, our calculated neutron pairing energies  vanish for the classic neutron shell closures N=20 and N=28 for calcium and argon isotopes. This is due to the weakening of  pairing correlations in the case of a magic nucleus. Also, the neutron pairing energy is exactly zero at N=32 and N=40. So, the $E_{pair}$ allows us to reproduce the classical neutron magic numbers N=20 and 28 and confirms the appearance of new ones: N=32 and 40.\\
			From the presented results in Figure~\ref{Epair}, we can observe  clearly  a sharp peak  at N =14 indicating a neutron shell closure for both nuclei under investigation. Closed shells at N=16 and 26 are always weak. 
%%%%%%%%%%%%%%%%%%%%%%%%%%%%%%%%%%%%%%%%%%%%%%%%%%%%%%%%%%%%%%%%%%%%%%%	
			\begin{figure}[h!]
				\begin{minipage}[b]{0.45\linewidth}
					\centering \includegraphics[scale=0.5]{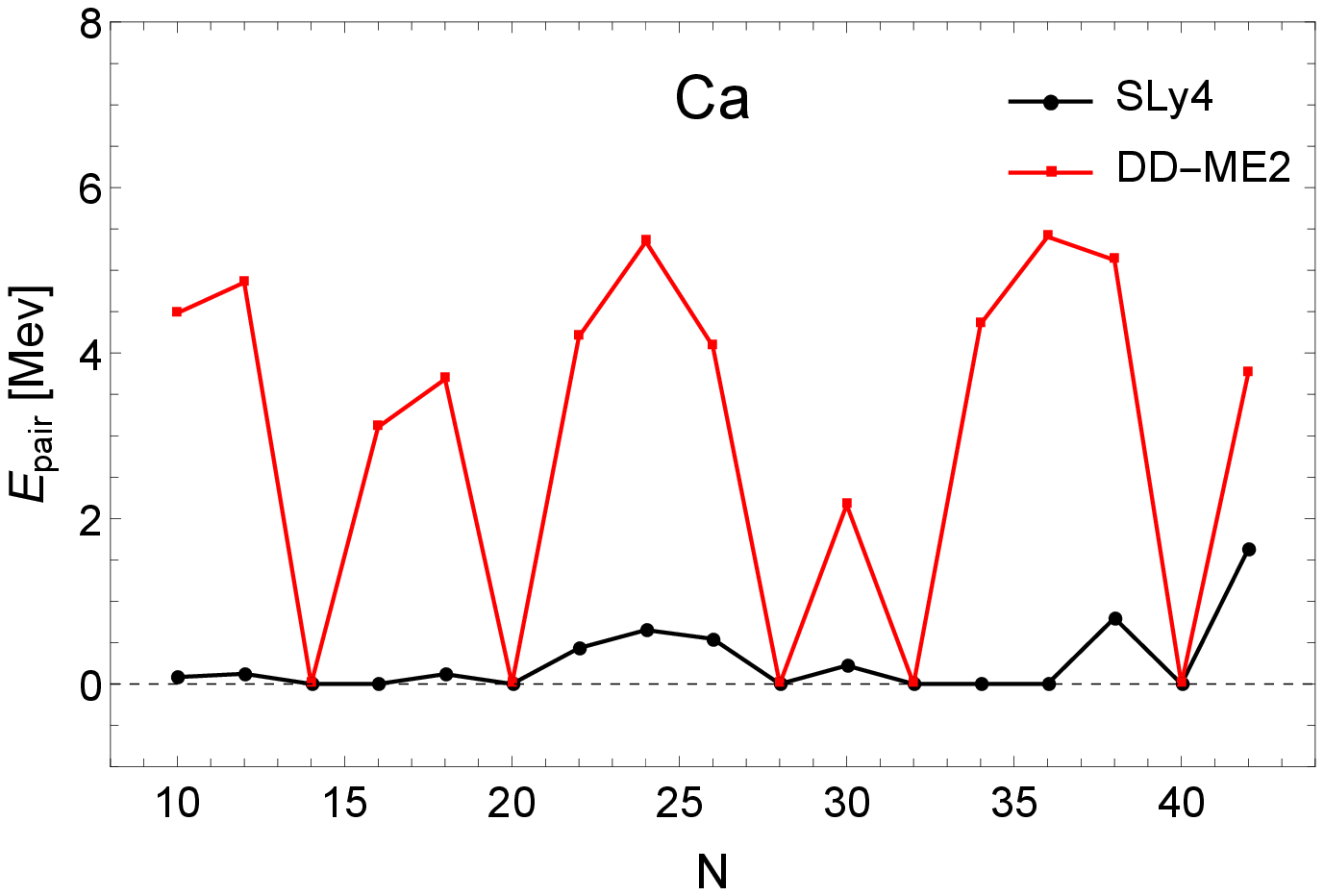}
					%  \caption{}
				\end{minipage}\hfill
				\begin{minipage}[b]{0.45\linewidth}
					\centering \includegraphics[scale=0.5]{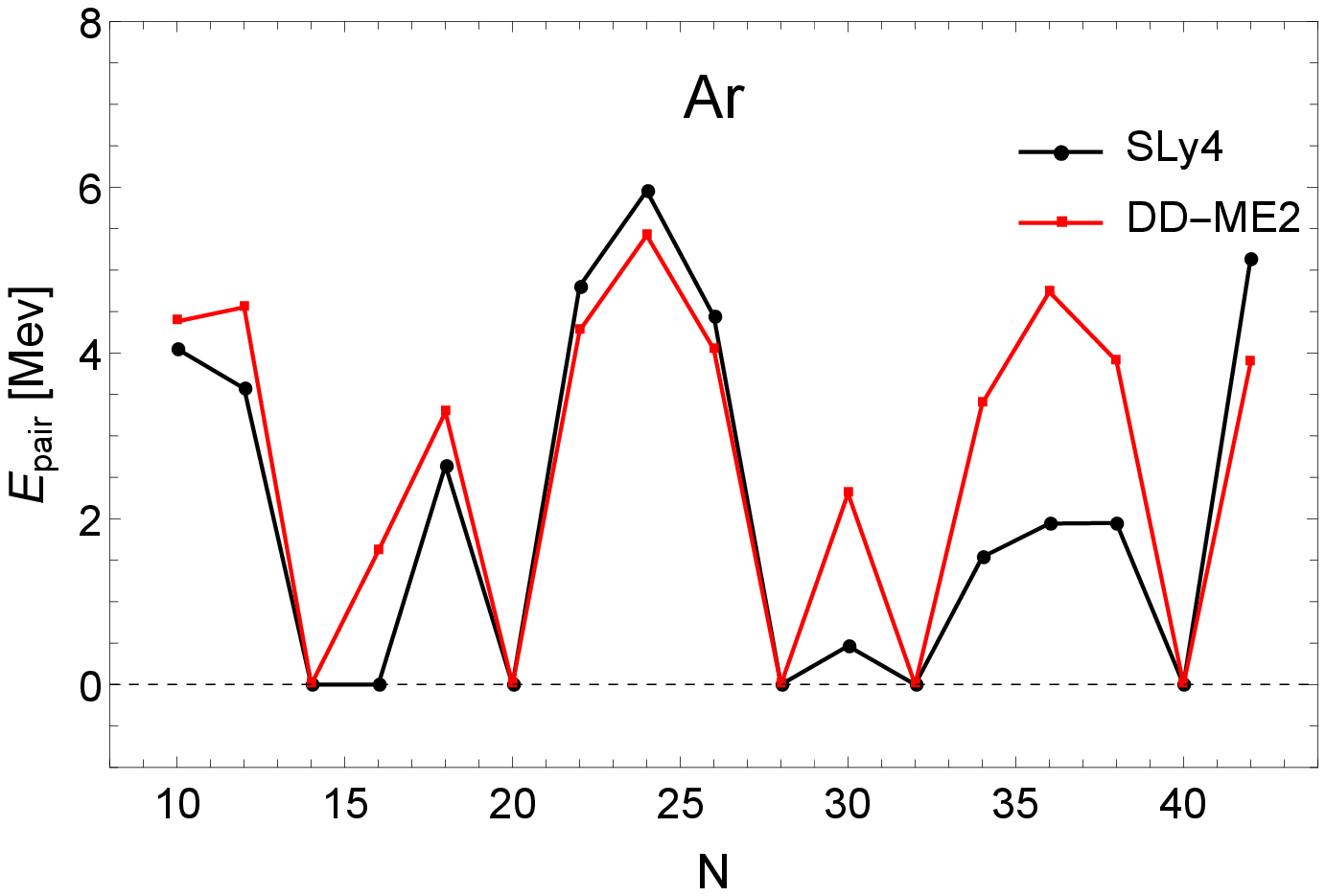}
					%  \caption{}
				\end{minipage}\hfill  
				\caption{(Color online) Calculated values of pairing energy as a function of N.}\label{Epair}
			\end{figure}
%%%%%%%%%%%%%%%%%%%%%%%%%%%%%%%%%%%%%%%%%%%%%%%%%%%%%%%%%%%%%%%%%%%%%%%	
			\subsection{Effective pairing gap and pairing tensor.}\label{Eff-pairing-gap}
			In HFBTHO calculations, the effective pairing gap defined as the mean value of the pairing field is given by
			\begin{equation}
			\bar{\Delta}=\frac{Tr(\Delta\rho)}{Tr(\rho)}
			\end{equation}
			with $\rho$ is the normal one-body density matrix and $\Delta$ is the pairing field.
			Such a defined effective gap has the same behavior as the spectral gaps in Ref \cite{Bender}. 
			While, in CDFT calculations, the computer code we have used gives information about pairing effect via the pairing tensor $\kappa$ which is given by eq(\ref{kappa}). 
			
			Both Effective pairing gap and pairing tensor are exactly zero for closed shell nuclei and their adjacent number.
			As shown in Figure~\ref{Gap}, it is seen that when approaching the classic neutron magic numbers  N=20 and N=28, the values of $\bar{\Delta}$ and $\kappa$ vanish. 
			A similar effect is seen around the neutron number N=32 and N=40, which correspond to new closed shells. \\
			Also, from the results  displayed in  Figure~\ref{Gap},   it is quite apparent that a shell closure is appearing at N=14 for both calcium and argon. Around N=16 and N=26, the values of $\bar{\Delta}$ are not zero, so there are no closed shells in this region.
%%%%%%%%%%%%%%%%%%%%%%%%%%%%%%%%%%%%%%%%%%%%%%%%%%%%%%%%%%%%%%%%%%%%%%%%
			\begin{figure}[h!]
				\begin{minipage}[b]{0.45\linewidth}
					\centering \includegraphics[scale=0.5]{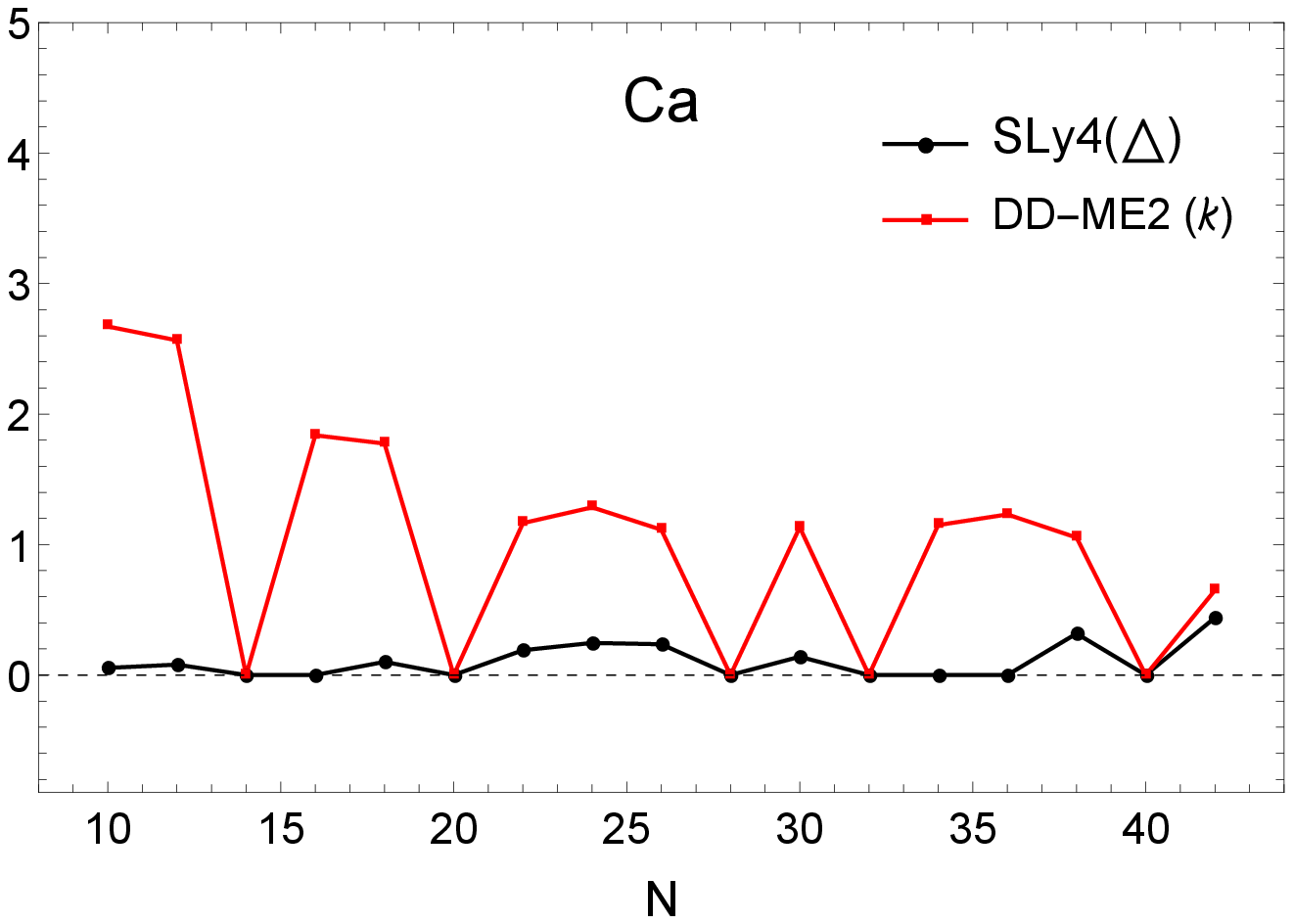}
					%  \caption{}
				\end{minipage}\hfill
				\begin{minipage}[b]{0.45\linewidth}
					\centering \includegraphics[scale=0.5]{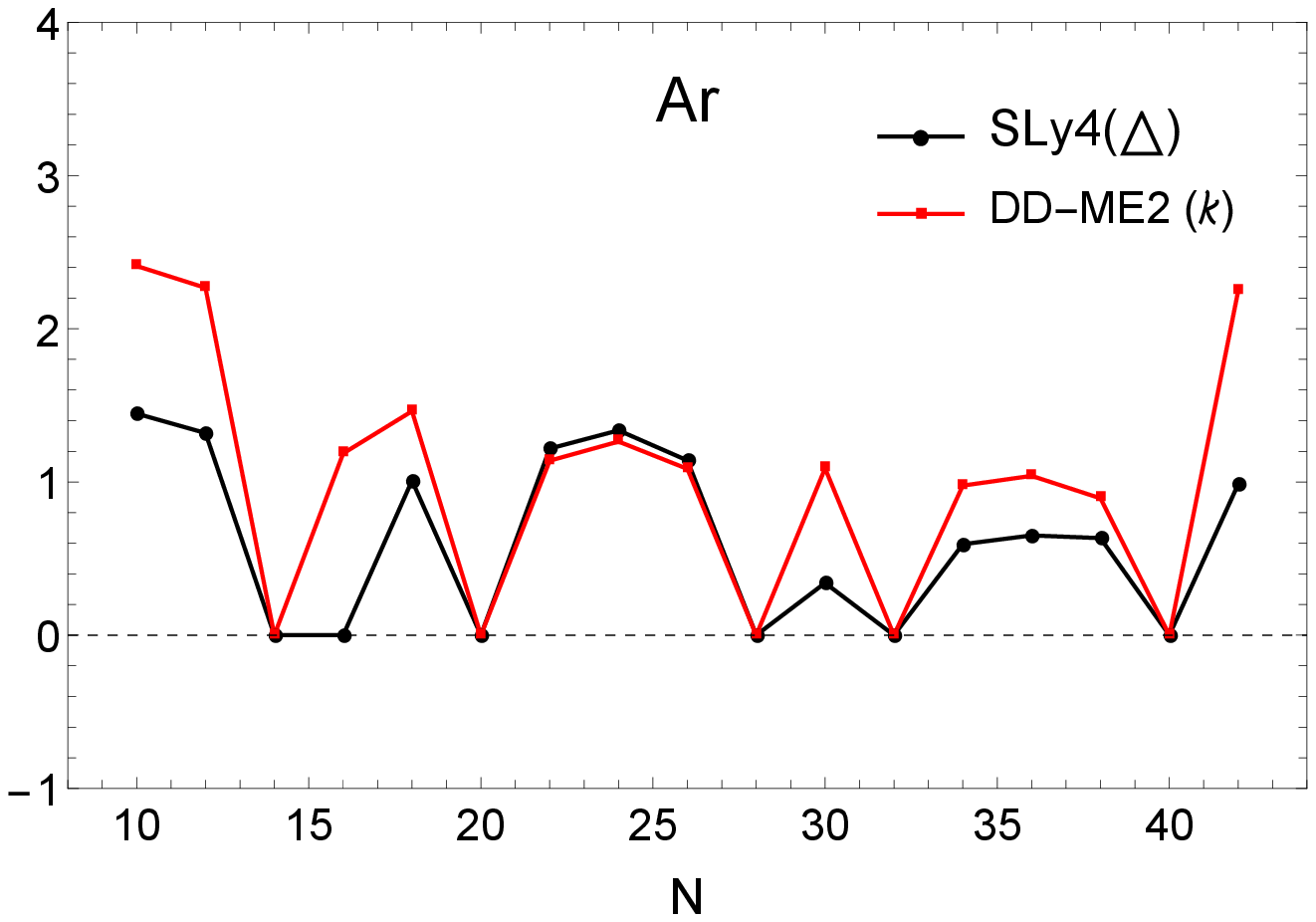}
					%  \caption{}
				\end{minipage}\hfill  
				\caption{Effective pairing gap $\bar{\Delta}$ and pairing tensor $\kappa$ of the even-even calcium (right) and argon (left) isotopes calculated with the HFB and CDFT models, respectively.}\label{Gap}
			\end{figure}
%%%%%%%%%%%%%%%%%%%%%%%%%%%%%%%%%%%%%%%%%%%%%%%%%%%%%%%%%%%%%%%%%%%%%%%	
\subsection{Single-particle spectrum}
The single-particle levels around the Fermi surface of the $_{20}Ca$ and $_{18}Ar$ nuclei, shown in Figure~\ref{fig:sp} have been calculated with CDFT+DD-ME2. It can be seen, that the gaps between the neutron $1p_{1/2}$ and $1d_{5/2}$, $1d_{3/2}$ and $1f_{7/2}$, and $1p_{1/2}$ and $1d_{5/2}$ states are significant enough to make a shell closure at N=8, N=20 and N=20, respectively. Also, the size of the N=14 ($\nu1d_{5/2}-\nu2s_{1/2}$), N=32 ($\nu2p_{3/2}-\nu1f_{5/2}$) and N=40 ($\nu2p_{1/2}-\nu1g_{9/2}$) gaps, which is almost similar for both nuclei $_{20}Ca$ and $_{18}Ar$, provides a similar behavior at classic magic numbers (20 and 28). Such a result corroborates our previously obtained ones in Secs. \ref{StwoN}, \ref{SoneN}, \ref{Shellgap}, \ref{Pairinggap}, \ref{Pairing energy} and \ref{Eff-pairing-gap} for the magic numbers N=32 and N=40.
% We conclude that 
\\
%%%%%%%%%%%%%%%%%%%%%%%%%%%%%%%%%%%%%%%%%%%%%%%%%%%%%%%%%%%%%%%%%%%%%%%	

\begin{figure}[h!]
	\begin{minipage}[b]{0.45\linewidth}
		\centering \includegraphics[scale=0.5]{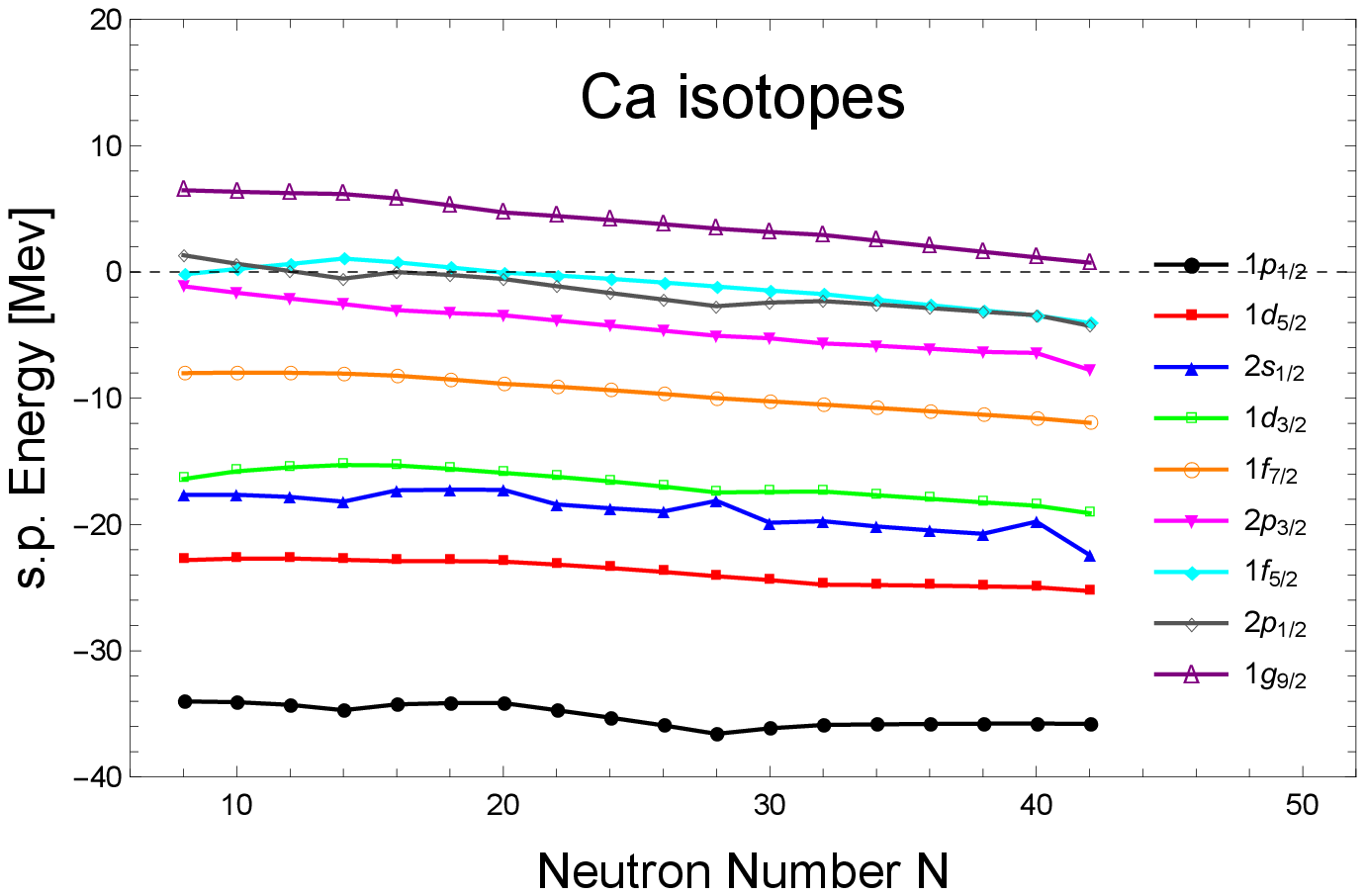}
		%  \caption{}
	\end{minipage}\hfill
	\begin{minipage}[b]{0.45\linewidth}
		\centering \includegraphics[scale=0.5]{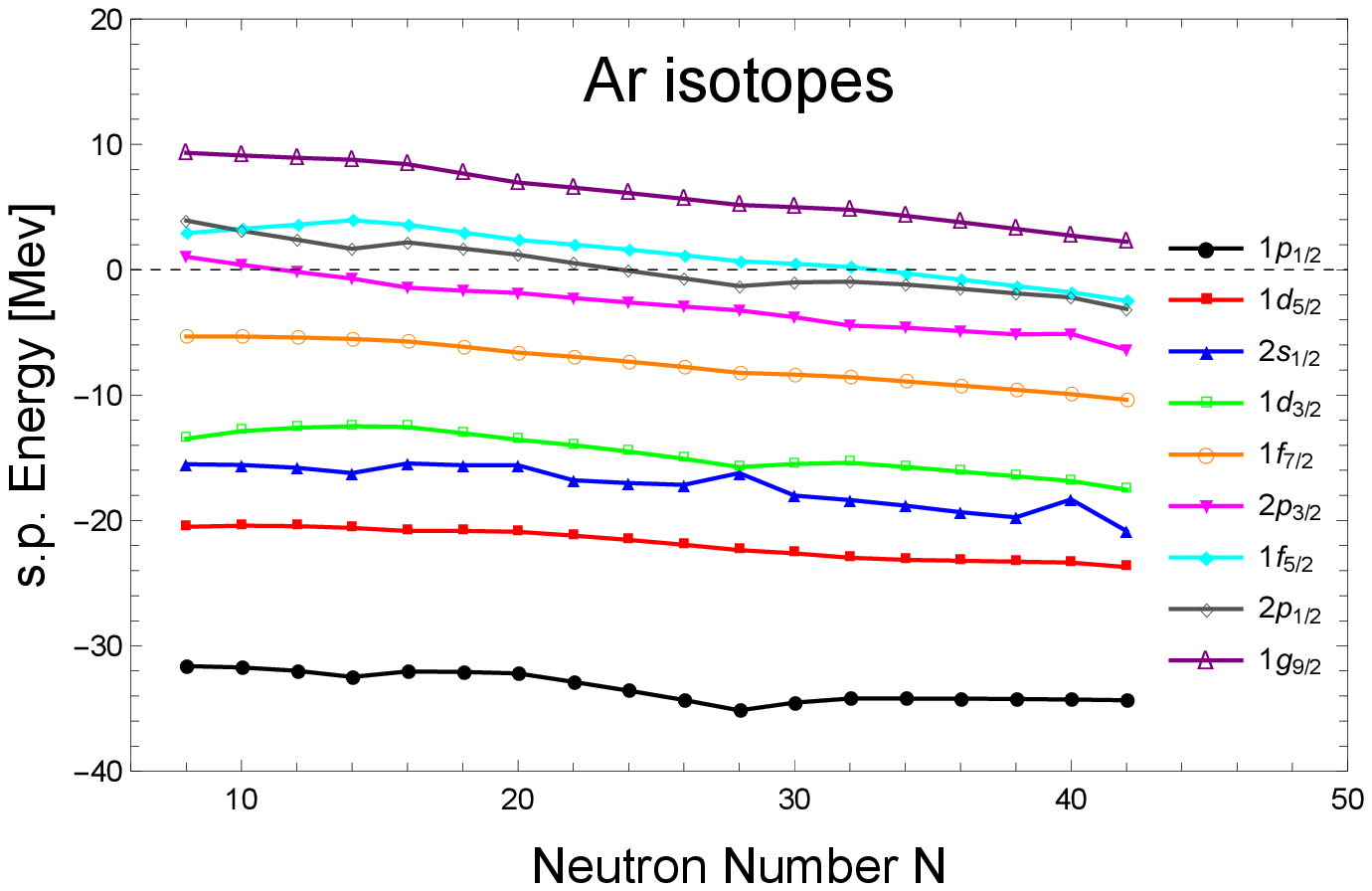}
		%  \caption{}
	\end{minipage}\hfill
	\caption{The single-particle spectrum as a function of neutron number N for  Ca isotopes (left panel) and Ar isotopes (right panel) within the CDFT calculations with DD-ME2.}\label{fig:spEV8}
\end{figure}
%%%%%%%%%%%%%%%%%%%%%%%%%%%%%%%%%%%%%%%%%%%%%%%%%%%%%%%%%%%%%%%%%%%%%%%%
As the single-particle spectrum is sensitive to the tensor force we redid the calculation by the EV8 code that takes into account the effect of this force. As shown in Figure \ref{fig:spEV8} The results obtained confirmed the magicity of the Ar and Ca nuclei at N = 32 and N = 40.
%%%%%%%%%%%%%%%%%%%%%%%%%%%%%%%%%%%%%%%%%%%%%%%%%%%%%%%%%%%%%%%%%%%%%%%	
\begin{figure}[h!]
	\begin{minipage}[b]{0.45\linewidth}
		\centering \includegraphics[scale=0.5]{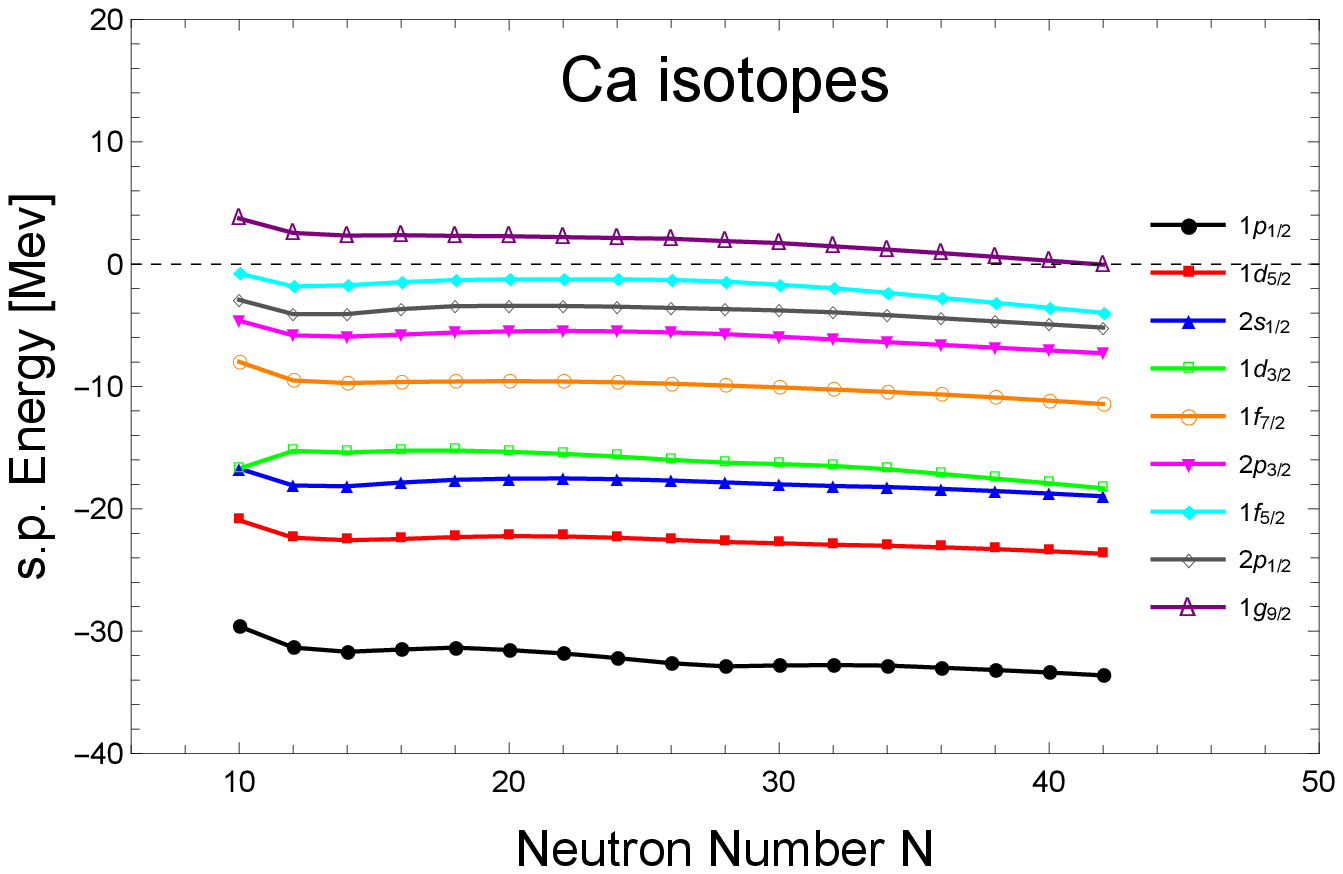}
		%  \caption{}
	\end{minipage}\hfill
	\begin{minipage}[b]{0.45\linewidth}
		\centering \includegraphics[scale=0.5]{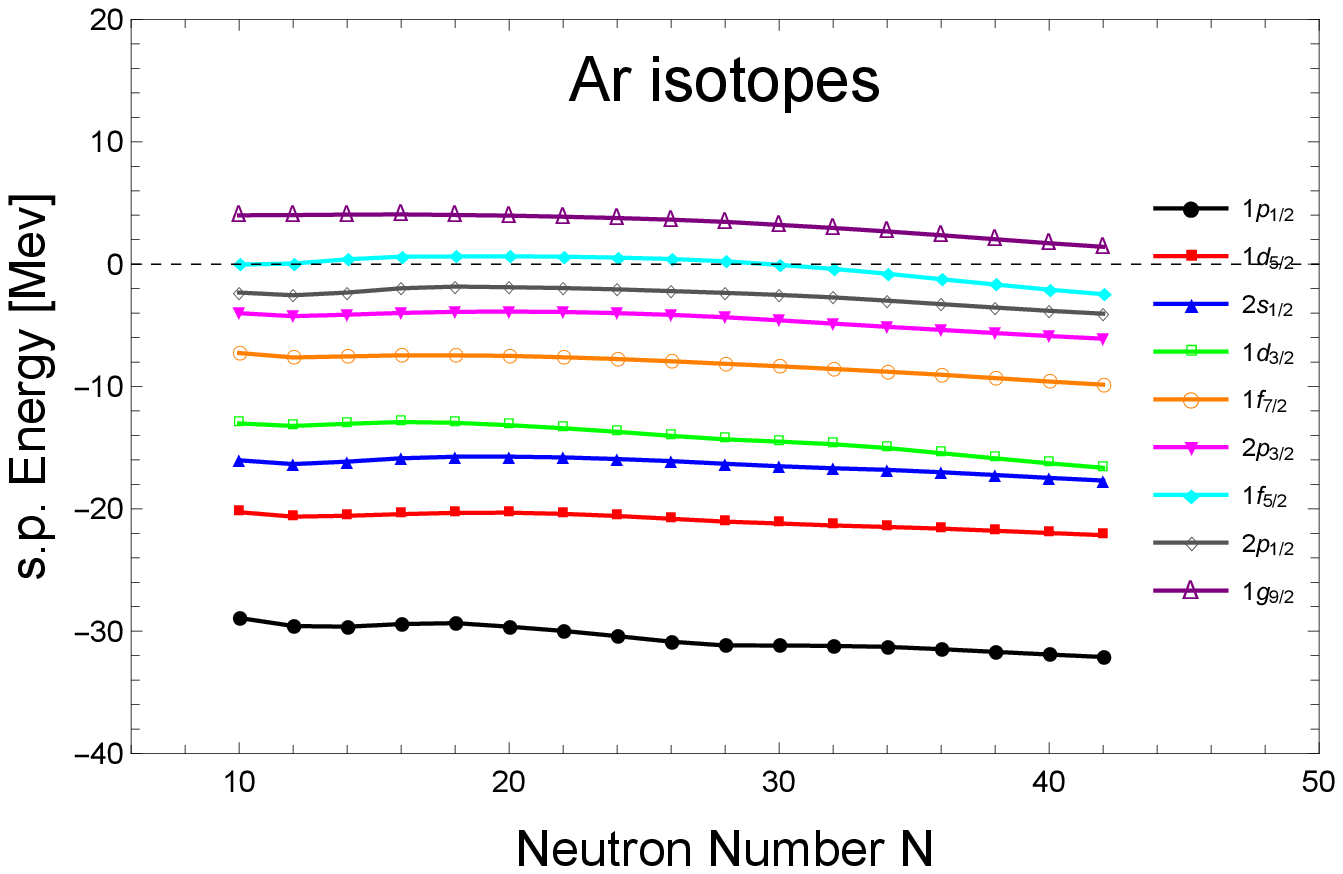}
		%  \caption{}
	\end{minipage}\hfill
	\caption{The single-particle spectrum as a function of neutron number N for  Ca isotopes (left panel) and Ar isotopes (right panel) within the tensor force.}\label{fig:sp}
\end{figure}
%%%%%%%%%%%%%%%%%%%%%%%%%%%%%%%%%%%%%%%%%%%%%%%%%%%%%%%%%%%%%%%%%%%%%%%%
			\section{Conclusion}
			In summary, the appearance of new shell closures in neutron-rich nuclei is studied by using two different approaches: the non relativistic Hartree-Fock-Bogoliubov (HFB) approach with SLy4 Skyrme force and a new generalized formula for pairing strength $V_{0}^{p,n}$ for both protons and neutrons, and the covariant density functional theory (CDFT) by using the DD-ME2 force. The obtained results  reproduced nicely the available experimental data. Based on the one- and two-neutron separation energies, two-neutron shell gap, neutron pairing gap, neutron pairing energy, effective pairing gap and pairing tensor, we reproduced the classic magic numbers N=20 and N=28   and  predicted that N=32 and N=40 are new ones in argon and calcium isotopes. In addition, as was obtained in Ref \cite{Janssens}, the shell closure at N=14 observed in oxygen was confirmed for calcium and argon in our investigation. Moreover, in the case of argon isotopes, N = 20 was not found to be a magic number neither in the experience nor in the FRDM predictions. This is, probably, due to the inversion of the standard sd-shell configuration and pf-shell intruder configuration as it has been proved in  $^{32}Mg$. However, the microscopic theories suggest N = 20 as a magic number. It seems rather that these theories  are failing in this respect. This observation paves the way for a future investigation.
					
			\section*{Acknowledgements}
			Discussions with T. Nik\v{s}i\'{c} from University of Zagreb, Faculty of Science, Physics Department, Croatia, are gratefully acknowledged.
			.


\section*{References}
			\begin{thebibliography}{99}
				\bibitem{Pud}B. S. Pudliner, V. R. Pandharipande, J. Carlson, R. B. Wiringa, Physical review letters, 74 (1995) 4396.
				\bibitem{Pudl}B. S. Pudliner, V. R. Pandharipande, J. Carlson, S. C. Pieper, R. B. Wiringa, Physical Review C, 56 (1997) 1720.
				\bibitem{Bender}M. Bender, P. H. Heenen and P. G. Reinhard, Rev. Mod. Phys. 75, 121 (2003).
				\bibitem{Serot} B.D. Serot, Rep. Prog. Phys. 55, 1855 (1992).
				\bibitem{Ring}P. Ring, Prog. Part. Nucl. Phys. 37, 193 (1996).
				\bibitem{Poschl}W. P\"{o}schl, D. Vretenar, G.A. Lalazissis and P. Ring, Phys. Rev. Lett. 79, 3841 (1997).
				\bibitem{Dobaczewsk1984} J. Dobaczewski, H. Flocard and J. Treiner, Nucl. Phys A422, 103 (1984).
				\bibitem{Davies}K.T.R. Davies, K.R.S. Devi, S.E. Koonin and M.R. Strayer, Vol. 3, in Treatise on Heavy Ion Science, edited by D. Bromley, volume 3, page 3, Plenum, New York, 1985.
				\bibitem{Negele}J.W. Negele, Rev. Mod. Phys. 54, 913 (1982).
				\bibitem{Carlson}Carlson et al., Rev. Mod. Phys. 87, 1067 (2015).
				\bibitem{Hagen}Hagen et al., Phys. Scr. 91, 063006 (2016).
				\bibitem{Barrett}Barrett et al., Prog Part.Nucl.Phys.69,131 (2013).
				\bibitem{Drut}Drut et al., Prog. Part.Nucl.Phys. 64, 120 (2010).
				\bibitem{Grasso}Grasso, Prog. Part. Nucl. Phys. 106, 256 (2019).
				\bibitem{Ozawa} A. Ozawa, T. Kobayashi, T. Suzuki, K. Yoshida, I. Tanihata, Phys. Rev. Lett. 84, 5493 (2000).
				\bibitem{tendeur} F. Tondeur, Proceedings of the 4th International Conference on Nuclear Far from Stability, Helsingør, Denmark (Cern, Geneva,1981), pp. 81–89.
				\bibitem{Huck}A. Huck, G. Klotz, A. Knipper, C. Miehé, C. Richard-Serre, G. Walter, A. Poves, H. L. Ravn, and G. Marguier, Phys. Rev. C 31, 2226 (1985).
				\bibitem{Kanungo} Kanungo, Phys. Scr. T152 (2013) 014002.
				\bibitem{Grawe}Grawe et al., Rep.Prog.Phys.70 (2007) 1525.
				\bibitem{Jurado} B. Jurado, Phys. Lett. B 649 (2007) 43.
				\bibitem{Nakamura} Nakamura et al. Prog.Part.Nucl.Phys.97 (2017)53.
				\bibitem{Sorlin} Sorlin, Phys. Scr. T152(2013)014003.
				\bibitem{Otsuka} Otsuka, Phys.Scr.T152(2013)014007.
				\bibitem{Ringp}P. Ring and P. Schuck, editor, The Nuclear Many-Body Problem, Springer Verlag, New York, 1980.
				\bibitem{chabanat}E. Chabanat, P. Bonche, P. Haensel, J. Meyer and R. Schaeffer, Nucl. Phys. A 635 (1998) 231-256.
				\bibitem{Landau} L. D. Landau, Sov. Phys. JETP 8, 70 (1959).
				\bibitem{Migdal}A. B. Migdal, editor, Theory of finite Fermi systems and Applications to atomic nuclei, Wilei Interscience, New York, 1967.
				\bibitem{Stoitsov}M. V. Stoitsov, J. Dobaczewski, W. Nazarewicz and P. Ring, Comp. Phys. Commun. 167, 43 (2005).
				\bibitem{Greiner}W. Greiner, J. A. Maruhn,{\it Nuclear Models} Berlin, Springer-Verlag, (1995).
				\bibitem{Chabanat}E. Chabanat, P. Bonche, P. Haensel, J. Meyer, R. Schaeffer, Nucl. Phys. A 635 (1998) 231-256.
				\bibitem{Abusara} H. Abusara, A. V. Afanasjev, and P. Ring, Phys. Rev. C85, 024314 (2012).
				\bibitem{Lalazissis} G. A. Lalazissis, T. Niksic, D. Vretenar, and P. Ring, Phys. Rev. C71, 024312 (2005).
				\bibitem{Niksic} T. Nik\v{s}i\'{c}, D. Vretenar, and P. Ring, Phys. Rev. C78, 034318 (2008).
				\bibitem{Niksik2014} T. Nik\v{s}i\'{c},  N. Paar, D. Vretenar et P. Ring, Computer Physics Communications 185 (2014) 1808-1821.
				\bibitem{Goeppert} M. Goeppert Mayer and J. H. Jensen, Elementary Theory of Nuclear Shell Structure (John Wiley and Sons, New York, 1955).
				\bibitem{Kucharek}H. Kucharek and P. Ring, Z. Phys. A 339, 23 (1991).
				\bibitem{Afanasjev}A. V. Afanasjev, P. Ring, and J. Konig, Nucl. Phys. A676, 196 (2000). 
				\bibitem{Dobaszewski1995}J. Dobaczewski, W. Nazarewicz, and T. R. Werner, Phys. Scr. T 56, 15 (1995).
				\bibitem{Fayanas}S. A. Fayans, S. V. Tolokonnikov, E. L. Trybov, and D. Zawischa, Phys. Lett. B 338, 1 (1994).
				\bibitem{Stoitsov2013} M. V. Stoitsov, N. Schunck, M. Kortelainen, N. Michel, H. Nam, E. Olsen, S. Wild, Comp. Phys. Commun. 184, 1592 (2013).
				\bibitem{Xia} Xia et al., At. Data Nucl. Data Tables 121 (2018),1
				\bibitem{Okolowicz} Okolowicz et al., Phys. Rep. 374 (2003) 271
				\bibitem{Forssen} Forssen et al, Phys. Scr. T152, 014022 (2013)
				\bibitem{Moller}P. M\"{o}ller, J.R. Nix, W.D. Myers, and W.J. Swiatecki. Nuclear ground-state masses and deformations. Atomic Data and Nuclear Data Tables, 59 (1995) 185-381.
				\bibitem{Elbassem}Y. EL Bassem and M. Oulne. Int. J. Mod. Phys. E 24 (2015) 1550073.
				\bibitem{FRDM}P. M\"{o}ller, A. J. Sierka, T. Ichikawab and H. Sagawa, {\it Atomic Data and Nuclear Data Tables} 109, 1-204 (2016).
				\bibitem{Exp} M. Wang, G. Audi, A. H. Wapstra and al. -{\it The Ame2012 atomic mass evaluation}. Chinese Physics C, 36, 1603 (2012).
				\bibitem{Chaudhuri} Chaudhuri et al., Phys. Rev. C 88, 054317 (2013).
				\bibitem{Yu} Yu. S. Lutostansky, S. M. Lukyanov, Yu. E. Penionzhkevich, M. V. Zverev. Neutron Drip Line in the Region of O-Mg Isotopes.No. JINR--6-115-2002. Joint of Institute for Nuclear Research, 2002.
				
							
			\end{thebibliography}
		\end{document}